\documentclass[longauth]{aaEC}

\usepackage{graphicx}
\usepackage{natbib}

\usepackage[table]{xcolor}

\usepackage{txfonts}
\usepackage[pdfencoding=auto,psdextra]{hyperref}
\hypersetup{
    colorlinks=true,
    linkcolor=blue,
    filecolor=magenta,      
    urlcolor=blue,
    citecolor=blue
}
\makeatletter
\renewcommand*\aa@pageof{, page \thepage{} of \pageref*{LastPage}}
\makeatother

\usepackage[utf8]{inputenc}

\usepackage[switch, modulo]{lineno}

\usepackage{euclid}

\usepackage[para,online,flushleft]{threeparttable}
\graphicspath{{./}{figures/}}

\newcommand{\papernumber}{}

\begin{document}
%
%
   \title{\Euclid Preparation. \papernumber The Cosmic Dawn Survey (DAWN) of the Euclid Deep and Auxiliary Fields
}

\newcommand{\orcid}[1]{} 
\author{Euclid Collaboration: C.~J.~R.~McPartland\orcid{0000-0003-0639-025X}\thanks{\email{conor.mcpartland@nbi.ku.dk}}\inst{\ref{aff1},\ref{aff2},\ref{aff3},\ref{aff4},\ref{aff5}}
\and L.~Zalesky\orcid{0000-0001-5680-2326}\inst{\ref{aff3}}
\and J.~R.~Weaver\orcid{0000-0003-1614-196X}\inst{\ref{aff6}}
\and S.~Toft\orcid{0000-0003-3631-7176}\inst{\ref{aff1},\ref{aff2}}
\and D.~B.~Sanders\orcid{0000-0002-1233-9998}\inst{\ref{aff3}}
\and B.~Mobasher\orcid{0000-0001-5846-4404}\inst{\ref{aff4}}
\and N.~Suzuki\orcid{0000-0001-7266-930X}\inst{\ref{aff7}}
\and I.~Szapudi\inst{\ref{aff3}}
\and I.~Valdes\inst{\ref{aff3}}
\and G.~Murphree\orcid{0009-0007-7266-8914}\inst{\ref{aff3}}
\and N.~Chartab\orcid{0000-0003-3691-937X}\inst{\ref{aff8}}
\and N.~Allen\orcid{0000-0001-9610-7950}\inst{\ref{aff1}}
\and S.~Taamoli\orcid{0000-0003-0749-4667}\inst{\ref{aff4}}
\and P.~R.~M.~Eisenhardt\inst{\ref{aff9}}
\and S.~Arnouts\inst{\ref{aff10}}
\and H.~Atek\orcid{0000-0002-7570-0824}\inst{\ref{aff11}}
\and J.~Brinchmann\orcid{0000-0003-4359-8797}\inst{\ref{aff12}}
\and M.~Castellano\orcid{0000-0001-9875-8263}\inst{\ref{aff13}}
\and R.~Chary\inst{\ref{aff14}}
\and O.~Ch\'{a}vez~Ortiz\orcid{0000-0003-2332-5505}\inst{\ref{aff15}}
\and J.-G.~Cuby\orcid{0000-0002-8767-1442}\inst{\ref{aff16},\ref{aff10}}
\and S.~L.~Finkelstein\orcid{0000-0001-8519-1130}\inst{\ref{aff15}}
\and T.~Goto\orcid{0000-0002-6821-8669}\inst{\ref{aff17}}
\and S.~Gwyn\orcid{0000-0001-8221-8406}\inst{\ref{aff18}}
\and Y.~Harikane\orcid{0000-0002-6047-430X}\inst{\ref{aff19}}
\and A.~K.~Inoue\orcid{0000-0002-7779-8677}\inst{\ref{aff20},\ref{aff21}}
\and H.~J.~McCracken\orcid{0000-0002-9489-7765}\inst{\ref{aff11}}
\and J.~J.~Mohr\orcid{0000-0002-6875-2087}\inst{\ref{aff22},\ref{aff23}}
\and P.~A.~Oesch\orcid{0000-0001-5851-6649}\inst{\ref{aff24},\ref{aff2},\ref{aff1}}
\and M.~Ouchi\orcid{0000-0002-1049-6658}\inst{\ref{aff25},\ref{aff19},\ref{aff26},\ref{aff27}}
\and M.~Oguri\inst{\ref{aff17},\ref{aff28}}
\and J.~Rhodes\inst{\ref{aff9}}
\and H.~J.~A.~Rottgering\orcid{0000-0001-8887-2257}\inst{\ref{aff29}}
\and M.~Sawicki\orcid{0000-0002-7712-7857}\inst{\ref{aff30}}
\and R.~Scaramella\orcid{0000-0003-2229-193X}\inst{\ref{aff13},\ref{aff31}}
\and C.~Scarlata\orcid{0000-0002-9136-8876}\inst{\ref{aff32}}
\and J.~D.~Silverman\orcid{0000-0002-0000-6977}\inst{\ref{aff27},\ref{aff33},\ref{aff34}}
\and D.~Stern\orcid{0000-0003-2686-9241}\inst{\ref{aff9}}
\and H.~I.~Teplitz\orcid{0000-0002-7064-5424}\inst{\ref{aff14}}
\and M.~Shuntov\orcid{0000-0002-7087-0701}\inst{\ref{aff35},\ref{aff5},\ref{aff2}}
\and B.~Altieri\orcid{0000-0003-3936-0284}\inst{\ref{aff36}}
\and A.~Amara\inst{\ref{aff37}}
\and S.~Andreon\orcid{0000-0002-2041-8784}\inst{\ref{aff38}}
\and N.~Auricchio\orcid{0000-0003-4444-8651}\inst{\ref{aff39}}
\and H.~Aussel\orcid{0000-0002-1371-5705}\inst{\ref{aff40}}
\and C.~Baccigalupi\orcid{0000-0002-8211-1630}\inst{\ref{aff41},\ref{aff42},\ref{aff43},\ref{aff44}}
\and M.~Baldi\orcid{0000-0003-4145-1943}\inst{\ref{aff45},\ref{aff39},\ref{aff46}}
\and S.~Bardelli\orcid{0000-0002-8900-0298}\inst{\ref{aff39}}
\and R.~Bender\orcid{0000-0001-7179-0626}\inst{\ref{aff23},\ref{aff22}}
\and D.~Bonino\orcid{0000-0002-3336-9977}\inst{\ref{aff47}}
\and E.~Branchini\orcid{0000-0002-0808-6908}\inst{\ref{aff48},\ref{aff49},\ref{aff38}}
\and M.~Brescia\orcid{0000-0001-9506-5680}\inst{\ref{aff50},\ref{aff51},\ref{aff52}}
\and S.~Camera\orcid{0000-0003-3399-3574}\inst{\ref{aff53},\ref{aff54},\ref{aff47}}
\and V.~Capobianco\orcid{0000-0002-3309-7692}\inst{\ref{aff47}}
\and C.~Carbone\orcid{0000-0003-0125-3563}\inst{\ref{aff55}}
\and J.~Carretero\orcid{0000-0002-3130-0204}\inst{\ref{aff56},\ref{aff57}}
\and S.~Casas\orcid{0000-0002-4751-5138}\inst{\ref{aff58}}
\and F.~J.~Castander\orcid{0000-0001-7316-4573}\inst{\ref{aff59},\ref{aff60}}
\and G.~Castignani\orcid{0000-0001-6831-0687}\inst{\ref{aff61},\ref{aff39}}
\and S.~Cavuoti\orcid{0000-0002-3787-4196}\inst{\ref{aff51},\ref{aff52}}
\and A.~Cimatti\inst{\ref{aff62}}
\and C.~Colodro-Conde\inst{\ref{aff63}}
\and G.~Congedo\orcid{0000-0003-2508-0046}\inst{\ref{aff64}}
\and C.~J.~Conselice\orcid{0000-0003-1949-7638}\inst{\ref{aff65}}
\and L.~Conversi\orcid{0000-0002-6710-8476}\inst{\ref{aff66},\ref{aff36}}
\and Y.~Copin\orcid{0000-0002-5317-7518}\inst{\ref{aff67}}
\and F.~Courbin\orcid{0000-0003-0758-6510}\inst{\ref{aff68}}
\and H.~M.~Courtois\orcid{0000-0003-0509-1776}\inst{\ref{aff69}}
\and A.~Da~Silva\orcid{0000-0002-6385-1609}\inst{\ref{aff70},\ref{aff71}}
\and H.~Degaudenzi\orcid{0000-0002-5887-6799}\inst{\ref{aff24}}
\and G.~De~Lucia\orcid{0000-0002-6220-9104}\inst{\ref{aff42}}
\and A.~M.~Di~Giorgio\orcid{0000-0002-4767-2360}\inst{\ref{aff72}}
\and J.~Dinis\inst{\ref{aff70},\ref{aff71}}
\and M.~Douspis\inst{\ref{aff73}}
\and F.~Dubath\orcid{0000-0002-6533-2810}\inst{\ref{aff24}}
\and X.~Dupac\inst{\ref{aff36}}
\and S.~Dusini\orcid{0000-0002-1128-0664}\inst{\ref{aff74}}
\and M.~Fabricius\inst{\ref{aff23},\ref{aff22}}
\and M.~Farina\orcid{0000-0002-3089-7846}\inst{\ref{aff72}}
\and S.~Farrens\orcid{0000-0002-9594-9387}\inst{\ref{aff40}}
\and S.~Ferriol\inst{\ref{aff67}}
\and S.~Fotopoulou\orcid{0000-0002-9686-254X}\inst{\ref{aff75}}
\and M.~Frailis\orcid{0000-0002-7400-2135}\inst{\ref{aff42}}
\and E.~Franceschi\orcid{0000-0002-0585-6591}\inst{\ref{aff39}}
\and M.~Fumana\orcid{0000-0001-6787-5950}\inst{\ref{aff55}}
\and S.~Galeotta\orcid{0000-0002-3748-5115}\inst{\ref{aff42}}
\and B.~Garilli\orcid{0000-0001-7455-8750}\inst{\ref{aff55}}
\and K.~George\orcid{0000-0002-1734-8455}\inst{\ref{aff22}}
\and B.~Gillis\orcid{0000-0002-4478-1270}\inst{\ref{aff64}}
\and C.~Giocoli\orcid{0000-0002-9590-7961}\inst{\ref{aff39},\ref{aff76}}
\and A.~Grazian\orcid{0000-0002-5688-0663}\inst{\ref{aff77}}
\and F.~Grupp\inst{\ref{aff23},\ref{aff22}}
\and L.~Guzzo\orcid{0000-0001-8264-5192}\inst{\ref{aff78},\ref{aff38}}
\and H.~Hoekstra\orcid{0000-0002-0641-3231}\inst{\ref{aff29}}
\and W.~Holmes\inst{\ref{aff9}}
\and I.~Hook\orcid{0000-0002-2960-978X}\inst{\ref{aff79}}
\and F.~Hormuth\inst{\ref{aff80}}
\and A.~Hornstrup\orcid{0000-0002-3363-0936}\inst{\ref{aff81},\ref{aff5}}
\and P.~Hudelot\inst{\ref{aff11}}
\and K.~Jahnke\orcid{0000-0003-3804-2137}\inst{\ref{aff82}}
\and E.~Keih\"anen\orcid{0000-0003-1804-7715}\inst{\ref{aff83}}
\and S.~Kermiche\orcid{0000-0002-0302-5735}\inst{\ref{aff84}}
\and A.~Kiessling\orcid{0000-0002-2590-1273}\inst{\ref{aff9}}
\and M.~Kilbinger\orcid{0000-0001-9513-7138}\inst{\ref{aff40}}
\and T.~Kitching\orcid{0000-0002-4061-4598}\inst{\ref{aff85}}
\and B.~Kubik\inst{\ref{aff67}}
\and M.~Kunz\orcid{0000-0002-3052-7394}\inst{\ref{aff86}}
\and H.~Kurki-Suonio\orcid{0000-0002-4618-3063}\inst{\ref{aff87},\ref{aff88}}
\and P.~B.~Lilje\orcid{0000-0003-4324-7794}\inst{\ref{aff89}}
\and V.~Lindholm\orcid{0000-0003-2317-5471}\inst{\ref{aff87},\ref{aff88}}
\and I.~Lloro\inst{\ref{aff90}}
\and G.~Mainetti\inst{\ref{aff91}}
\and E.~Maiorano\orcid{0000-0003-2593-4355}\inst{\ref{aff39}}
\and O.~Mansutti\orcid{0000-0001-5758-4658}\inst{\ref{aff42}}
\and O.~Marggraf\orcid{0000-0001-7242-3852}\inst{\ref{aff92}}
\and K.~Markovic\orcid{0000-0001-6764-073X}\inst{\ref{aff9}}
\and M.~Martinelli\orcid{0000-0002-6943-7732}\inst{\ref{aff13},\ref{aff31}}
\and N.~Martinet\orcid{0000-0003-2786-7790}\inst{\ref{aff10}}
\and F.~Marulli\orcid{0000-0002-8850-0303}\inst{\ref{aff61},\ref{aff39},\ref{aff46}}
\and R.~Massey\orcid{0000-0002-6085-3780}\inst{\ref{aff93}}
\and S.~Maurogordato\inst{\ref{aff94}}
\and E.~Medinaceli\orcid{0000-0002-4040-7783}\inst{\ref{aff39}}
\and S.~Mei\orcid{0000-0002-2849-559X}\inst{\ref{aff95}}
\and M.~Melchior\inst{\ref{aff96}}
\and Y.~Mellier\inst{\ref{aff35},\ref{aff11}}
\and M.~Meneghetti\orcid{0000-0003-1225-7084}\inst{\ref{aff39},\ref{aff46}}
\and E.~Merlin\orcid{0000-0001-6870-8900}\inst{\ref{aff13}}
\and G.~Meylan\inst{\ref{aff68}}
\and M.~Moresco\orcid{0000-0002-7616-7136}\inst{\ref{aff61},\ref{aff39}}
\and L.~Moscardini\orcid{0000-0002-3473-6716}\inst{\ref{aff61},\ref{aff39},\ref{aff46}}
\and E.~Munari\orcid{0000-0002-1751-5946}\inst{\ref{aff42}}
\and R.~Nakajima\inst{\ref{aff92}}
\and C.~Neissner\orcid{0000-0001-8524-4968}\inst{\ref{aff56},\ref{aff57}}
\and S.-M.~Niemi\inst{\ref{aff97}}
\and J.~W.~Nightingale\orcid{0000-0002-8987-7401}\inst{\ref{aff98},\ref{aff93}}
\and C.~Padilla\orcid{0000-0001-7951-0166}\inst{\ref{aff56}}
\and S.~Paltani\orcid{0000-0002-8108-9179}\inst{\ref{aff24}}
\and F.~Pasian\orcid{0000-0002-4869-3227}\inst{\ref{aff42}}
\and K.~Pedersen\inst{\ref{aff99}}
\and W.~J.~Percival\orcid{0000-0002-0644-5727}\inst{\ref{aff100},\ref{aff101},\ref{aff102}}
\and V.~Pettorino\inst{\ref{aff97}}
\and G.~Polenta\orcid{0000-0003-4067-9196}\inst{\ref{aff103}}
\and M.~Poncet\inst{\ref{aff104}}
\and L.~A.~Popa\inst{\ref{aff105}}
\and L.~Pozzetti\orcid{0000-0001-7085-0412}\inst{\ref{aff39}}
\and F.~Raison\orcid{0000-0002-7819-6918}\inst{\ref{aff23}}
\and R.~Rebolo\inst{\ref{aff63},\ref{aff106}}
\and A.~Renzi\orcid{0000-0001-9856-1970}\inst{\ref{aff107},\ref{aff74}}
\and G.~Riccio\inst{\ref{aff51}}
\and E.~Romelli\orcid{0000-0003-3069-9222}\inst{\ref{aff42}}
\and M.~Roncarelli\orcid{0000-0001-9587-7822}\inst{\ref{aff39}}
\and E.~Rossetti\inst{\ref{aff45}}
\and R.~Saglia\orcid{0000-0003-0378-7032}\inst{\ref{aff22},\ref{aff23}}
\and Z.~Sakr\orcid{0000-0002-4823-3757}\inst{\ref{aff108},\ref{aff109},\ref{aff110}}
\and A.~G.~S\'anchez\orcid{0000-0003-1198-831X}\inst{\ref{aff23}}
\and D.~Sapone\orcid{0000-0001-7089-4503}\inst{\ref{aff111}}
\and B.~Sartoris\inst{\ref{aff22},\ref{aff42}}
\and M.~Schirmer\orcid{0000-0003-2568-9994}\inst{\ref{aff82}}
\and P.~Schneider\orcid{0000-0001-8561-2679}\inst{\ref{aff92}}
\and T.~Schrabback\orcid{0000-0002-6987-7834}\inst{\ref{aff112}}
\and A.~Secroun\orcid{0000-0003-0505-3710}\inst{\ref{aff84}}
\and G.~Seidel\orcid{0000-0003-2907-353X}\inst{\ref{aff82}}
\and S.~Serrano\orcid{0000-0002-0211-2861}\inst{\ref{aff60},\ref{aff59},\ref{aff113}}
\and C.~Sirignano\orcid{0000-0002-0995-7146}\inst{\ref{aff107},\ref{aff74}}
\and G.~Sirri\orcid{0000-0003-2626-2853}\inst{\ref{aff46}}
\and L.~Stanco\orcid{0000-0002-9706-5104}\inst{\ref{aff74}}
\and J.~Steinwagner\inst{\ref{aff23}}
\and C.~Surace\orcid{0000-0003-2592-0113}\inst{\ref{aff10}}
\and P.~Tallada-Crespi\orcid{0000-0002-1336-8328}\inst{\ref{aff114},\ref{aff57}}
\and D.~Tavagnacco\orcid{0000-0001-7475-9894}\inst{\ref{aff42}}
\and I.~Tereno\inst{\ref{aff70},\ref{aff115}}
\and R.~Toledo-Moreo\orcid{0000-0002-2997-4859}\inst{\ref{aff116}}
\and F.~Torradeflot\orcid{0000-0003-1160-1517}\inst{\ref{aff57},\ref{aff114}}
\and I.~Tutusaus\orcid{0000-0002-3199-0399}\inst{\ref{aff109}}
\and E.~A.~Valentijn\inst{\ref{aff117}}
\and L.~Valenziano\orcid{0000-0002-1170-0104}\inst{\ref{aff39},\ref{aff118}}
\and T.~Vassallo\orcid{0000-0001-6512-6358}\inst{\ref{aff22},\ref{aff42}}
\and A.~Veropalumbo\orcid{0000-0003-2387-1194}\inst{\ref{aff38},\ref{aff49}}
\and Y.~Wang\orcid{0000-0002-4749-2984}\inst{\ref{aff14}}
\and J.~Weller\orcid{0000-0002-8282-2010}\inst{\ref{aff22},\ref{aff23}}
\and G.~Zamorani\orcid{0000-0002-2318-301X}\inst{\ref{aff39}}
\and J.~Zoubian\inst{\ref{aff84}}
\and E.~Zucca\orcid{0000-0002-5845-8132}\inst{\ref{aff39}}
\and A.~Biviano\orcid{0000-0002-0857-0732}\inst{\ref{aff42},\ref{aff44}}
\and M.~Bolzonella\orcid{0000-0003-3278-4607}\inst{\ref{aff39}}
\and A.~Boucaud\orcid{0000-0001-7387-2633}\inst{\ref{aff95}}
\and E.~Bozzo\orcid{0000-0002-8201-1525}\inst{\ref{aff24}}
\and C.~Burigana\orcid{0000-0002-3005-5796}\inst{\ref{aff119},\ref{aff118}}
\and D.~Di~Ferdinando\inst{\ref{aff46}}
\and R.~Farinelli\inst{\ref{aff39}}
\and J.~Gracia-Carpio\inst{\ref{aff23}}
\and N.~Mauri\orcid{0000-0001-8196-1548}\inst{\ref{aff62},\ref{aff46}}
\and V.~Scottez\inst{\ref{aff35},\ref{aff120}}
\and M.~Tenti\orcid{0000-0002-4254-5901}\inst{\ref{aff46}}
\and M.~Viel\orcid{0000-0002-2642-5707}\inst{\ref{aff44},\ref{aff42},\ref{aff41},\ref{aff43},\ref{aff121}}
\and M.~Wiesmann\orcid{0009-0000-8199-5860}\inst{\ref{aff89}}
\and Y.~Akrami\orcid{0000-0002-2407-7956}\inst{\ref{aff122},\ref{aff123}}
\and V.~Allevato\orcid{0000-0001-7232-5152}\inst{\ref{aff51}}
\and S.~Anselmi\orcid{0000-0002-3579-9583}\inst{\ref{aff74},\ref{aff107},\ref{aff124}}
\and M.~Ballardini\orcid{0000-0003-4481-3559}\inst{\ref{aff125},\ref{aff39},\ref{aff126}}
\and M.~Bethermin\orcid{0000-0002-3915-2015}\inst{\ref{aff127},\ref{aff10}}
\and S.~Borgani\orcid{0000-0001-6151-6439}\inst{\ref{aff128},\ref{aff44},\ref{aff42},\ref{aff43}}
\and A.~S.~Borlaff\orcid{0000-0003-3249-4431}\inst{\ref{aff129},\ref{aff130}}
\and S.~Bruton\orcid{0000-0002-6503-5218}\inst{\ref{aff32}}
\and R.~Cabanac\orcid{0000-0001-6679-2600}\inst{\ref{aff109}}
\and A.~Calabro\orcid{0000-0003-2536-1614}\inst{\ref{aff13}}
\and G.~Ca\~{n}as-Herrera\orcid{0000-0003-2796-2149}\inst{\ref{aff97},\ref{aff131}}
\and A.~Cappi\inst{\ref{aff39},\ref{aff94}}
\and C.~S.~Carvalho\inst{\ref{aff115}}
\and T.~Castro\orcid{0000-0002-6292-3228}\inst{\ref{aff42},\ref{aff43},\ref{aff44},\ref{aff121}}
\and K.~C.~Chambers\orcid{0000-0001-6965-7789}\inst{\ref{aff3}}
\and S.~Contarini\orcid{0000-0002-9843-723X}\inst{\ref{aff23},\ref{aff61}}
\and A.~R.~Cooray\orcid{0000-0002-3892-0190}\inst{\ref{aff132}}
\and J.~Coupon\inst{\ref{aff24}}
\and S.~Davini\orcid{0000-0003-3269-1718}\inst{\ref{aff49}}
\and S.~de~la~Torre\inst{\ref{aff10}}
\and G.~Desprez\inst{\ref{aff30}}
\and A.~D\'iaz-S\'anchez\orcid{0000-0003-0748-4768}\inst{\ref{aff133}}
\and S.~Di~Domizio\orcid{0000-0003-2863-5895}\inst{\ref{aff48},\ref{aff49}}
\and H.~Dole\orcid{0000-0002-9767-3839}\inst{\ref{aff73}}
\and J.~A.~Escartin~Vigo\inst{\ref{aff23}}
\and S.~Escoffier\orcid{0000-0002-2847-7498}\inst{\ref{aff84}}
\and A.~G.~Ferrari\orcid{0009-0005-5266-4110}\inst{\ref{aff62},\ref{aff46}}
\and P.~G.~Ferreira\inst{\ref{aff134}}
\and I.~Ferrero\orcid{0000-0002-1295-1132}\inst{\ref{aff89}}
\and F.~Finelli\orcid{0000-0002-6694-3269}\inst{\ref{aff39},\ref{aff118}}
\and F.~Fornari\orcid{0000-0003-2979-6738}\inst{\ref{aff118}}
\and L.~Gabarra\orcid{0000-0002-8486-8856}\inst{\ref{aff134}}
\and K.~Ganga\orcid{0000-0001-8159-8208}\inst{\ref{aff95}}
\and J.~Garc\'ia-Bellido\orcid{0000-0002-9370-8360}\inst{\ref{aff122}}
\and V.~Gautard\inst{\ref{aff135}}
\and E.~Gaztanaga\orcid{0000-0001-9632-0815}\inst{\ref{aff59},\ref{aff60},\ref{aff136}}
\and F.~Giacomini\orcid{0000-0002-3129-2814}\inst{\ref{aff46}}
\and G.~Gozaliasl\orcid{0000-0002-0236-919X}\inst{\ref{aff137},\ref{aff87}}
\and A.~Gregorio\orcid{0000-0003-4028-8785}\inst{\ref{aff128},\ref{aff42},\ref{aff43}}
\and A.~Hall\orcid{0000-0002-3139-8651}\inst{\ref{aff64}}
\and W.~G.~Hartley\inst{\ref{aff24}}
\and H.~Hildebrandt\orcid{0000-0002-9814-3338}\inst{\ref{aff138}}
\and J.~Hjorth\orcid{0000-0002-4571-2306}\inst{\ref{aff139}}
\and M.~Huertas-Company\orcid{0000-0002-1416-8483}\inst{\ref{aff63},\ref{aff140},\ref{aff141},\ref{aff142}}
\and O.~Ilbert\orcid{0000-0002-7303-4397}\inst{\ref{aff10}}
\and J.~J.~E.~Kajava\orcid{0000-0002-3010-8333}\inst{\ref{aff143},\ref{aff144}}
\and V.~Kansal\orcid{0000-0002-4008-6078}\inst{\ref{aff145},\ref{aff146}}
\and D.~Karagiannis\orcid{0000-0002-4927-0816}\inst{\ref{aff147},\ref{aff148}}
\and C.~C.~Kirkpatrick\inst{\ref{aff83}}
\and L.~Legrand\orcid{0000-0003-0610-5252}\inst{\ref{aff149}}
\and G.~Libet\inst{\ref{aff104}}
\and A.~Loureiro\orcid{0000-0002-4371-0876}\inst{\ref{aff150},\ref{aff151}}
\and J.~Macias-Perez\orcid{0000-0002-5385-2763}\inst{\ref{aff152}}
\and G.~Maggio\orcid{0000-0003-4020-4836}\inst{\ref{aff42}}
\and M.~Magliocchetti\orcid{0000-0001-9158-4838}\inst{\ref{aff72}}
\and C.~Mancini\orcid{0000-0002-4297-0561}\inst{\ref{aff55}}
\and F.~Mannucci\orcid{0000-0002-4803-2381}\inst{\ref{aff153}}
\and R.~Maoli\orcid{0000-0002-6065-3025}\inst{\ref{aff154},\ref{aff13}}
\and C.~J.~A.~P.~Martins\orcid{0000-0002-4886-9261}\inst{\ref{aff155},\ref{aff12}}
\and S.~Matthew\inst{\ref{aff64}}
\and M.~Maturi\orcid{0000-0002-3517-2422}\inst{\ref{aff108},\ref{aff156}}
\and L.~Maurin\orcid{0000-0002-8406-0857}\inst{\ref{aff73}}
\and R.~B.~Metcalf\orcid{0000-0003-3167-2574}\inst{\ref{aff61},\ref{aff39}}
\and P.~Monaco\orcid{0000-0003-2083-7564}\inst{\ref{aff128},\ref{aff42},\ref{aff43},\ref{aff44}}
\and C.~Moretti\orcid{0000-0003-3314-8936}\inst{\ref{aff41},\ref{aff121},\ref{aff42},\ref{aff44},\ref{aff43}}
\and G.~Morgante\inst{\ref{aff39}}
\and P.~Musi\inst{\ref{aff157}}
\and Nicholas~A.~Walton\orcid{0000-0003-3983-8778}\inst{\ref{aff158}}
\and J.~Odier\orcid{0000-0002-1650-2246}\inst{\ref{aff152}}
\and L.~Patrizii\inst{\ref{aff46}}
\and M.~P{\"o}ntinen\orcid{0000-0001-5442-2530}\inst{\ref{aff87}}
\and V.~Popa\inst{\ref{aff105}}
\and C.~Porciani\orcid{0000-0002-7797-2508}\inst{\ref{aff92}}
\and D.~Potter\orcid{0000-0002-0757-5195}\inst{\ref{aff159}}
\and P.~Reimberg\orcid{0000-0003-3410-0280}\inst{\ref{aff35}}
\and I.~Risso\orcid{0000-0003-2525-7761}\inst{\ref{aff160}}
\and P.-F.~Rocci\inst{\ref{aff73}}
\and M.~Sahl\'en\orcid{0000-0003-0973-4804}\inst{\ref{aff161}}
\and A.~Schneider\orcid{0000-0001-7055-8104}\inst{\ref{aff159}}
\and M.~Sereno\orcid{0000-0003-0302-0325}\inst{\ref{aff39},\ref{aff46}}
\and P.~Simon\inst{\ref{aff92}}
\and A.~Spurio~Mancini\orcid{0000-0001-5698-0990}\inst{\ref{aff85}}
\and S.~A.~Stanford\orcid{0000-0003-0122-0841}\inst{\ref{aff162}}
\and C.~Tao\orcid{0000-0001-7961-8177}\inst{\ref{aff84}}
\and G.~Testera\inst{\ref{aff49}}
\and R.~Teyssier\orcid{0000-0001-7689-0933}\inst{\ref{aff163}}
\and S.~Tosi\orcid{0000-0002-7275-9193}\inst{\ref{aff48},\ref{aff38},\ref{aff49}}
\and A.~Troja\orcid{0000-0003-0239-4595}\inst{\ref{aff107},\ref{aff74}}
\and M.~Tucci\inst{\ref{aff24}}
\and C.~Valieri\inst{\ref{aff46}}
\and J.~Valiviita\orcid{0000-0001-6225-3693}\inst{\ref{aff87},\ref{aff88}}
\and D.~Vergani\orcid{0000-0003-0898-2216}\inst{\ref{aff39}}
\and G.~Verza\orcid{0000-0002-1886-8348}\inst{\ref{aff164},\ref{aff165}}
\and F.~Shankar\orcid{0000-0001-8973-5051}\inst{\ref{aff166}}}
										   
\institute{Cosmic Dawn Center (DAWN)\label{aff1}
\and
Niels Bohr Institute, University of Copenhagen, Jagtvej 128, 2200 Copenhagen, Denmark\label{aff2}
\and
Institute for Astronomy, University of Hawaii, 2680 Woodlawn Drive, Honolulu, HI 96822, USA\label{aff3}
\and
Physics and Astronomy Department, University of California, 900 University Ave., Riverside, CA 92521, USA\label{aff4}
\and
Cosmic Dawn Center (DAWN), Denmark\label{aff5}
\and
Department of Astronomy, University of Massachusetts, Amherst, MA 01003, USA\label{aff6}
\and
Lawrence Berkeley National Laboratory, One Cyclotron Road, Berkeley, CA 94720, USA\label{aff7}
\and
Carnegie Observatories, Pasadena, CA 91101, USA\label{aff8}
\and
Jet Propulsion Laboratory, California Institute of Technology, 4800 Oak Grove Drive, Pasadena, CA, 91109, USA\label{aff9}
\and
Aix-Marseille Universit\'e, CNRS, CNES, LAM, Marseille, France\label{aff10}
\and
Institut d'Astrophysique de Paris, UMR 7095, CNRS, and Sorbonne Universit\'e, 98 bis boulevard Arago, 75014 Paris, France\label{aff11}
\and
Instituto de Astrof\'isica e Ci\^encias do Espa\c{c}o, Universidade do Porto, CAUP, Rua das Estrelas, PT4150-762 Porto, Portugal\label{aff12}
\and
INAF-Osservatorio Astronomico di Roma, Via Frascati 33, 00078 Monteporzio Catone, Italy\label{aff13}
\and
Infrared Processing and Analysis Center, California Institute of Technology, Pasadena, CA 91125, USA\label{aff14}
\and
The University of Texas at Austin, Austin, TX, 78712, USA\label{aff15}
\and
Canada-France-Hawaii Telescope, 65-1238 Mamalahoa Hwy, Kamuela, HI 96743, USA\label{aff16}
\and
Center for Frontier Science, Chiba University, 1-33 Yayoi-cho, Inage-ku, Chiba 263-8522, Japan\label{aff17}
\and
NRC Herzberg, 5071 West Saanich Rd, Victoria, BC V9E 2E7, Canada\label{aff18}
\and
Institute for Cosmic Ray Research, The University of Tokyo, 5-1-5 Kashiwanoha, Kashiwa, Chiba 277-8582, Japan\label{aff19}
\and
Department of Physics, School of Advanced Science and Engineering, Faculty of Science and Engineering, Waseda University, 3-4-1 Okubo, Shinjuku, 169-8555 Tokyo, Japan\label{aff20}
\and
Waseda Research Institute for Science and Engineering, Faculty of Science and Engineering, Waseda University, 3-4-1 Okubo, Shinjuku, Tokyo 169-8555, Japan\label{aff21}
\and
Universit\"ats-Sternwarte M\"unchen, Fakult\"at f\"ur Physik, Ludwig-Maximilians-Universit\"at M\"unchen, Scheinerstrasse 1, 81679 M\"unchen, Germany\label{aff22}
\and
Max Planck Institute for Extraterrestrial Physics, Giessenbachstr. 1, 85748 Garching, Germany\label{aff23}
\and
Department of Astronomy, University of Geneva, ch. d'Ecogia 16, 1290 Versoix, Switzerland\label{aff24}
\and
National Astronomical Observatory of Japan, 2-21-1 Osawa, Mitaka, Tokyo 181-8588, Japan\label{aff25}
\and
Department of Astronomical Science, SOKENDAI (The Graduate University for Advanced Studies), Osawa 2-21-1, Mitaka, Tokyo, 181-8588, Japan\label{aff26}
\and
Kavli Institute for the Physics and Mathematics of the Universe (WPI), University of Tokyo, Kashiwa, Chiba 277-8583, Japan\label{aff27}
\and
Department of Physics, Graduate School of Science, Chiba University, 1-33 Yayoi-Cho, Inage-Ku, Chiba 263-8522, Japan\label{aff28}
\and
Leiden Observatory, Leiden University, Einsteinweg 55, 2333 CC Leiden, The Netherlands\label{aff29}
\and
Department of Astronomy \& Physics and Institute for Computational Astrophysics, Saint Mary's University, 923 Robie Street, Halifax, Nova Scotia, B3H 3C3, Canada\label{aff30}
\and
INFN-Sezione di Roma, Piazzale Aldo Moro, 2 - c/o Dipartimento di Fisica, Edificio G. Marconi, 00185 Roma, Italy\label{aff31}
\and
Minnesota Institute for Astrophysics, University of Minnesota, 116 Church St SE, Minneapolis, MN 55455, USA\label{aff32}
\and
Department of Astronomy, School of Science, The University of Tokyo, 7-3-1 Hongo, Bunkyo, Tokyo 113-0033, Japan\label{aff33}
\and
Center for Data-Driven Discovery, Kavli IPMU (WPI), UTIAS, The University of Tokyo, Kashiwa, Chiba 277-8583, Japan\label{aff34}
\and
Institut d'Astrophysique de Paris, 98bis Boulevard Arago, 75014, Paris, France\label{aff35}
\and
ESAC/ESA, Camino Bajo del Castillo, s/n., Urb. Villafranca del Castillo, 28692 Villanueva de la Ca\~nada, Madrid, Spain\label{aff36}
\and
School of Mathematics and Physics, University of Surrey, Guildford, Surrey, GU2 7XH, UK\label{aff37}
\and
INAF-Osservatorio Astronomico di Brera, Via Brera 28, 20122 Milano, Italy\label{aff38}
\and
INAF-Osservatorio di Astrofisica e Scienza dello Spazio di Bologna, Via Piero Gobetti 93/3, 40129 Bologna, Italy\label{aff39}
\and
Universit\'e Paris-Saclay, Universit\'e Paris Cit\'e, CEA, CNRS, AIM, 91191, Gif-sur-Yvette, France\label{aff40}
\and
SISSA, International School for Advanced Studies, Via Bonomea 265, 34136 Trieste TS, Italy\label{aff41}
\and
INAF-Osservatorio Astronomico di Trieste, Via G. B. Tiepolo 11, 34143 Trieste, Italy\label{aff42}
\and
INFN, Sezione di Trieste, Via Valerio 2, 34127 Trieste TS, Italy\label{aff43}
\and
IFPU, Institute for Fundamental Physics of the Universe, via Beirut 2, 34151 Trieste, Italy\label{aff44}
\and
Dipartimento di Fisica e Astronomia, Universit\`a di Bologna, Via Gobetti 93/2, 40129 Bologna, Italy\label{aff45}
\and
INFN-Sezione di Bologna, Viale Berti Pichat 6/2, 40127 Bologna, Italy\label{aff46}
\and
INAF-Osservatorio Astrofisico di Torino, Via Osservatorio 20, 10025 Pino Torinese (TO), Italy\label{aff47}
\and
Dipartimento di Fisica, Universit\`a di Genova, Via Dodecaneso 33, 16146, Genova, Italy\label{aff48}
\and
INFN-Sezione di Genova, Via Dodecaneso 33, 16146, Genova, Italy\label{aff49}
\and
Department of Physics "E. Pancini", University Federico II, Via Cinthia 6, 80126, Napoli, Italy\label{aff50}
\and
INAF-Osservatorio Astronomico di Capodimonte, Via Moiariello 16, 80131 Napoli, Italy\label{aff51}
\and
INFN section of Naples, Via Cinthia 6, 80126, Napoli, Italy\label{aff52}
\and
Dipartimento di Fisica, Universit\`a degli Studi di Torino, Via P. Giuria 1, 10125 Torino, Italy\label{aff53}
\and
INFN-Sezione di Torino, Via P. Giuria 1, 10125 Torino, Italy\label{aff54}
\and
INAF-IASF Milano, Via Alfonso Corti 12, 20133 Milano, Italy\label{aff55}
\and
Institut de F\'{i}sica d'Altes Energies (IFAE), The Barcelona Institute of Science and Technology, Campus UAB, 08193 Bellaterra (Barcelona), Spain\label{aff56}
\and
Port d'Informaci\'{o} Cient\'{i}fica, Campus UAB, C. Albareda s/n, 08193 Bellaterra (Barcelona), Spain\label{aff57}
\and
Institute for Theoretical Particle Physics and Cosmology (TTK), RWTH Aachen University, 52056 Aachen, Germany\label{aff58}
\and
Institute of Space Sciences (ICE, CSIC), Campus UAB, Carrer de Can Magrans, s/n, 08193 Barcelona, Spain\label{aff59}
\and
Institut d'Estudis Espacials de Catalunya (IEEC),  Edifici RDIT, Campus UPC, 08860 Castelldefels, Barcelona, Spain\label{aff60}
\and
Dipartimento di Fisica e Astronomia "Augusto Righi" - Alma Mater Studiorum Universit\`a di Bologna, via Piero Gobetti 93/2, 40129 Bologna, Italy\label{aff61}
\and
Dipartimento di Fisica e Astronomia "Augusto Righi" - Alma Mater Studiorum Universit\`a di Bologna, Viale Berti Pichat 6/2, 40127 Bologna, Italy\label{aff62}
\and
Instituto de Astrof\'isica de Canarias, Calle V\'ia L\'actea s/n, 38204, San Crist\'obal de La Laguna, Tenerife, Spain\label{aff63}
\and
Institute for Astronomy, University of Edinburgh, Royal Observatory, Blackford Hill, Edinburgh EH9 3HJ, UK\label{aff64}
\and
Jodrell Bank Centre for Astrophysics, Department of Physics and Astronomy, University of Manchester, Oxford Road, Manchester M13 9PL, UK\label{aff65}
\and
European Space Agency/ESRIN, Largo Galileo Galilei 1, 00044 Frascati, Roma, Italy\label{aff66}
\and
Universit\'e Claude Bernard Lyon 1, CNRS/IN2P3, IP2I Lyon, UMR 5822, Villeurbanne, F-69100, France\label{aff67}
\and
Institute of Physics, Laboratory of Astrophysics, Ecole Polytechnique F\'ed\'erale de Lausanne (EPFL), Observatoire de Sauverny, 1290 Versoix, Switzerland\label{aff68}
\and
UCB Lyon 1, CNRS/IN2P3, IUF, IP2I Lyon, 4 rue Enrico Fermi, 69622 Villeurbanne, France\label{aff69}
\and
Departamento de F\'isica, Faculdade de Ci\^encias, Universidade de Lisboa, Edif\'icio C8, Campo Grande, PT1749-016 Lisboa, Portugal\label{aff70}
\and
Instituto de Astrof\'isica e Ci\^encias do Espa\c{c}o, Faculdade de Ci\^encias, Universidade de Lisboa, Campo Grande, 1749-016 Lisboa, Portugal\label{aff71}
\and
INAF-Istituto di Astrofisica e Planetologia Spaziali, via del Fosso del Cavaliere, 100, 00100 Roma, Italy\label{aff72}
\and
Universit\'e Paris-Saclay, CNRS, Institut d'astrophysique spatiale, 91405, Orsay, France\label{aff73}
\and
INFN-Padova, Via Marzolo 8, 35131 Padova, Italy\label{aff74}
\and
School of Physics, HH Wills Physics Laboratory, University of Bristol, Tyndall Avenue, Bristol, BS8 1TL, UK\label{aff75}
\and
Istituto Nazionale di Fisica Nucleare, Sezione di Bologna, Via Irnerio 46, 40126 Bologna, Italy\label{aff76}
\and
INAF-Osservatorio Astronomico di Padova, Via dell'Osservatorio 5, 35122 Padova, Italy\label{aff77}
\and
Dipartimento di Fisica "Aldo Pontremoli", Universit\`a degli Studi di Milano, Via Celoria 16, 20133 Milano, Italy\label{aff78}
\and
Department of Physics, Lancaster University, Lancaster, LA1 4YB, UK\label{aff79}
\and
von Hoerner \& Sulger GmbH, Schlossplatz 8, 68723 Schwetzingen, Germany\label{aff80}
\and
Technical University of Denmark, Elektrovej 327, 2800 Kgs. Lyngby, Denmark\label{aff81}
\and
Max-Planck-Institut f\"ur Astronomie, K\"onigstuhl 17, 69117 Heidelberg, Germany\label{aff82}
\and
Department of Physics and Helsinki Institute of Physics, Gustaf H\"allstr\"omin katu 2, 00014 University of Helsinki, Finland\label{aff83}
\and
Aix-Marseille Universit\'e, CNRS/IN2P3, CPPM, Marseille, France\label{aff84}
\and
Mullard Space Science Laboratory, University College London, Holmbury St Mary, Dorking, Surrey RH5 6NT, UK\label{aff85}
\and
Universit\'e de Gen\`eve, D\'epartement de Physique Th\'eorique and Centre for Astroparticle Physics, 24 quai Ernest-Ansermet, CH-1211 Gen\`eve 4, Switzerland\label{aff86}
\and
Department of Physics, P.O. Box 64, 00014 University of Helsinki, Finland\label{aff87}
\and
Helsinki Institute of Physics, Gustaf H{\"a}llstr{\"o}min katu 2, University of Helsinki, Helsinki, Finland\label{aff88}
\and
Institute of Theoretical Astrophysics, University of Oslo, P.O. Box 1029 Blindern, 0315 Oslo, Norway\label{aff89}
\and
NOVA optical infrared instrumentation group at ASTRON, Oude Hoogeveensedijk 4, 7991PD, Dwingeloo, The Netherlands\label{aff90}
\and
Centre de Calcul de l'IN2P3/CNRS, 21 avenue Pierre de Coubertin 69627 Villeurbanne Cedex, France\label{aff91}
\and
Universit\"at Bonn, Argelander-Institut f\"ur Astronomie, Auf dem H\"ugel 71, 53121 Bonn, Germany\label{aff92}
\and
Department of Physics, Institute for Computational Cosmology, Durham University, South Road, DH1 3LE, UK\label{aff93}
\and
Universit\'e C\^{o}te d'Azur, Observatoire de la C\^{o}te d'Azur, CNRS, Laboratoire Lagrange, Bd de l'Observatoire, CS 34229, 06304 Nice cedex 4, France\label{aff94}
\and
Universit\'e Paris Cit\'e, CNRS, Astroparticule et Cosmologie, 75013 Paris, France\label{aff95}
\and
University of Applied Sciences and Arts of Northwestern Switzerland, School of Engineering, 5210 Windisch, Switzerland\label{aff96}
\and
European Space Agency/ESTEC, Keplerlaan 1, 2201 AZ Noordwijk, The Netherlands\label{aff97}
\and
School of Mathematics, Statistics and Physics, Newcastle University, Herschel Building, Newcastle-upon-Tyne, NE1 7RU, UK\label{aff98}
\and
Department of Physics and Astronomy, University of Aarhus, Ny Munkegade 120, DK-8000 Aarhus C, Denmark\label{aff99}
\and
Waterloo Centre for Astrophysics, University of Waterloo, Waterloo, Ontario N2L 3G1, Canada\label{aff100}
\and
Department of Physics and Astronomy, University of Waterloo, Waterloo, Ontario N2L 3G1, Canada\label{aff101}
\and
Perimeter Institute for Theoretical Physics, Waterloo, Ontario N2L 2Y5, Canada\label{aff102}
\and
Space Science Data Center, Italian Space Agency, via del Politecnico snc, 00133 Roma, Italy\label{aff103}
\and
Centre National d'Etudes Spatiales -- Centre spatial de Toulouse, 18 avenue Edouard Belin, 31401 Toulouse Cedex 9, France\label{aff104}
\and
Institute of Space Science, Str. Atomistilor, nr. 409 M\u{a}gurele, Ilfov, 077125, Romania\label{aff105}
\and
Departamento de Astrof\'isica, Universidad de La Laguna, 38206, La Laguna, Tenerife, Spain\label{aff106}
\and
Dipartimento di Fisica e Astronomia "G. Galilei", Universit\`a di Padova, Via Marzolo 8, 35131 Padova, Italy\label{aff107}
\and
Institut f\"ur Theoretische Physik, University of Heidelberg, Philosophenweg 16, 69120 Heidelberg, Germany\label{aff108}
\and
Institut de Recherche en Astrophysique et Plan\'etologie (IRAP), Universit\'e de Toulouse, CNRS, UPS, CNES, 14 Av. Edouard Belin, 31400 Toulouse, France\label{aff109}
\and
Universit\'e St Joseph; Faculty of Sciences, Beirut, Lebanon\label{aff110}
\and
Departamento de F\'isica, FCFM, Universidad de Chile, Blanco Encalada 2008, Santiago, Chile\label{aff111}
\and
Universit\"at Innsbruck, Institut f\"ur Astro- und Teilchenphysik, Technikerstr. 25/8, 6020 Innsbruck, Austria\label{aff112}
\and
Satlantis, University Science Park, Sede Bld 48940, Leioa-Bilbao, Spain\label{aff113}
\and
Centro de Investigaciones Energ\'eticas, Medioambientales y Tecnol\'ogicas (CIEMAT), Avenida Complutense 40, 28040 Madrid, Spain\label{aff114}
\and
Instituto de Astrof\'isica e Ci\^encias do Espa\c{c}o, Faculdade de Ci\^encias, Universidade de Lisboa, Tapada da Ajuda, 1349-018 Lisboa, Portugal\label{aff115}
\and
Universidad Polit\'ecnica de Cartagena, Departamento de Electr\'onica y Tecnolog\'ia de Computadoras,  Plaza del Hospital 1, 30202 Cartagena, Spain\label{aff116}
\and
Kapteyn Astronomical Institute, University of Groningen, PO Box 800, 9700 AV Groningen, The Netherlands\label{aff117}
\and
INFN-Bologna, Via Irnerio 46, 40126 Bologna, Italy\label{aff118}
\and
INAF, Istituto di Radioastronomia, Via Piero Gobetti 101, 40129 Bologna, Italy\label{aff119}
\and
Junia, EPA department, 41 Bd Vauban, 59800 Lille, France\label{aff120}
\and
ICSC - Centro Nazionale di Ricerca in High Performance Computing, Big Data e Quantum Computing, Via Magnanelli 2, Bologna, Italy\label{aff121}
\and
Instituto de F\'isica Te\'orica UAM-CSIC, Campus de Cantoblanco, 28049 Madrid, Spain\label{aff122}
\and
CERCA/ISO, Department of Physics, Case Western Reserve University, 10900 Euclid Avenue, Cleveland, OH 44106, USA\label{aff123}
\and
Laboratoire Univers et Th\'eorie, Observatoire de Paris, Universit\'e PSL, Universit\'e Paris Cit\'e, CNRS, 92190 Meudon, France\label{aff124}
\and
Dipartimento di Fisica e Scienze della Terra, Universit\`a degli Studi di Ferrara, Via Giuseppe Saragat 1, 44122 Ferrara, Italy\label{aff125}
\and
Istituto Nazionale di Fisica Nucleare, Sezione di Ferrara, Via Giuseppe Saragat 1, 44122 Ferrara, Italy\label{aff126}
\and
Universit\'e de Strasbourg, CNRS, Observatoire astronomique de Strasbourg, UMR 7550, 67000 Strasbourg, France\label{aff127}
\and
Dipartimento di Fisica - Sezione di Astronomia, Universit\`a di Trieste, Via Tiepolo 11, 34131 Trieste, Italy\label{aff128}
\and
NASA Ames Research Center, Moffett Field, CA 94035, USA\label{aff129}
\and
Bay Area Environmental Research Institute, Moffett Field, California 94035, USA\label{aff130}
\and
Institute Lorentz, Leiden University, Niels Bohrweg 2, 2333 CA Leiden, The Netherlands\label{aff131}
\and
Department of Physics \& Astronomy, University of California Irvine, Irvine CA 92697, USA\label{aff132}
\and
Departamento F\'isica Aplicada, Universidad Polit\'ecnica de Cartagena, Campus Muralla del Mar, 30202 Cartagena, Murcia, Spain\label{aff133}
\and
Department of Physics, Oxford University, Keble Road, Oxford OX1 3RH, UK\label{aff134}
\and
CEA Saclay, DFR/IRFU, Service d'Astrophysique, Bat. 709, 91191 Gif-sur-Yvette, France\label{aff135}
\and
Institute of Cosmology and Gravitation, University of Portsmouth, Portsmouth PO1 3FX, UK\label{aff136}
\and
Department of Computer Science, Aalto University, PO Box 15400, Espoo, FI-00 076, Finland\label{aff137}
\and
Ruhr University Bochum, Faculty of Physics and Astronomy, Astronomical Institute (AIRUB), German Centre for Cosmological Lensing (GCCL), 44780 Bochum, Germany\label{aff138}
\and
DARK, Niels Bohr Institute, University of Copenhagen, Jagtvej 155, 2200 Copenhagen, Denmark\label{aff139}
\and
Instituto de Astrof\'isica de Canarias (IAC); Departamento de Astrof\'isica, Universidad de La Laguna (ULL), 38200, La Laguna, Tenerife, Spain\label{aff140}
\and
Universit\'e PSL, Observatoire de Paris, Sorbonne Universit\'e, CNRS, LERMA, 75014, Paris, France\label{aff141}
\and
Universit\'e Paris-Cit\'e, 5 Rue Thomas Mann, 75013, Paris, France\label{aff142}
\and
Department of Physics and Astronomy, Vesilinnantie 5, 20014 University of Turku, Finland\label{aff143}
\and
Serco for European Space Agency (ESA), Camino bajo del Castillo, s/n, Urbanizacion Villafranca del Castillo, Villanueva de la Ca\~nada, 28692 Madrid, Spain\label{aff144}
\and
ARC Centre of Excellence for Dark Matter Particle Physics, Melbourne, Australia\label{aff145}
\and
Centre for Astrophysics \& Supercomputing, Swinburne University of Technology,  Hawthorn, Victoria 3122, Australia\label{aff146}
\and
School of Physics and Astronomy, Queen Mary University of London, Mile End Road, London E1 4NS, UK\label{aff147}
\and
Department of Physics and Astronomy, University of the Western Cape, Bellville, Cape Town, 7535, South Africa\label{aff148}
\and
ICTP South American Institute for Fundamental Research, Instituto de F\'{\i}sica Te\'orica, Universidade Estadual Paulista, S\~ao Paulo, Brazil\label{aff149}
\and
Oskar Klein Centre for Cosmoparticle Physics, Department of Physics, Stockholm University, Stockholm, SE-106 91, Sweden\label{aff150}
\and
Astrophysics Group, Blackett Laboratory, Imperial College London, London SW7 2AZ, UK\label{aff151}
\and
Univ. Grenoble Alpes, CNRS, Grenoble INP, LPSC-IN2P3, 53, Avenue des Martyrs, 38000, Grenoble, France\label{aff152}
\and
INAF-Osservatorio Astrofisico di Arcetri, Largo E. Fermi 5, 50125, Firenze, Italy\label{aff153}
\and
Dipartimento di Fisica, Sapienza Universit\`a di Roma, Piazzale Aldo Moro 2, 00185 Roma, Italy\label{aff154}
\and
Centro de Astrof\'{\i}sica da Universidade do Porto, Rua das Estrelas, 4150-762 Porto, Portugal\label{aff155}
\and
Zentrum f\"ur Astronomie, Universit\"at Heidelberg, Philosophenweg 12, 69120 Heidelberg, Germany\label{aff156}
\and
Thales Alenia Space -- Euclid satellite Prime contractor, Strada Antica di Collegno 253, 10146 Torino, Italy\label{aff157}
\and
Institute of Astronomy, University of Cambridge, Madingley Road, Cambridge CB3 0HA, UK\label{aff158}
\and
Department of Astrophysics, University of Zurich, Winterthurerstrasse 190, 8057 Zurich, Switzerland\label{aff159}
\and
Dipartimento di Fisica, Universit\`a degli studi di Genova, and INFN-Sezione di Genova, via Dodecaneso 33, 16146, Genova, Italy\label{aff160}
\and
Theoretical astrophysics, Department of Physics and Astronomy, Uppsala University, Box 515, 751 20 Uppsala, Sweden\label{aff161}
\and
Department of Physics and Astronomy, University of California, Davis, CA 95616, USA\label{aff162}
\and
Department of Astrophysical Sciences, Peyton Hall, Princeton University, Princeton, NJ 08544, USA\label{aff163}
\and
Center for Cosmology and Particle Physics, Department of Physics, New York University, New York, NY 10003, USA\label{aff164}
\and
Center for Computational Astrophysics, Flatiron Institute, 162 5th Avenue, 10010, New York, NY, USA\label{aff165}
\and
School of Physics \& Astronomy, University of Southampton, Highfield Campus, Southampton SO17 1BJ, UK\label{aff166}}    


%
%
 \abstract{
   \Euclid will provide deep NIR imaging to $\sim$26.5 AB magnitude over $\sim$59 deg$^2$ in its deep and auxiliary fields. The Cosmic DAWN survey complements the deep \Euclid data with matched depth multiwavelength imaging and spectroscopy in the UV--IR to provide consistently processed \Euclid selected photometric catalogs, accurate photometric redshifts, and measurements of galaxy properties to a redshift of $z\sim 10$. In this paper, we present an overview of the survey, including the footprints of the survey fields, the existing and planned observations, and the primary science goals for the combined data set.}
%
%
%
%
   \titlerunning{\Euclid Preparation. \papernumber. The DAWN Survey}
   \authorrunning{McPartland et al.}
   
   \maketitle
%
%
%
%
   
\section{\label{sc:Intro}Introduction}

The \Euclid mission \citep{laureijs_euclid_2011,EuclidOverview,VIS,NISP} is designed to constrain the properties of dark matter and dark energy through weak lensing, galaxy cluster counts, and clustering measurements. The majority of the six year mission will be spent carrying out a wide area imaging and spectroscopic survey \citep[the Euclid Wide Survey: EWS;][]{Scaramella22} covering roughly \num{14000}\,deg$^2$ of the extragalactic sky. The EWS will measure the shape and colour of billions of galaxies from imaging observations in a single broad visible band (\IE) and three near-infrared (NIR) bands (\YE, \JE, \HE) with expected depths of 25 AB (10$\sigma$ for extended sources) and 24.5 AB (5$\sigma$ for point sources) respectively. The spectroscopic component of EWS will measure redshifts for around thirty million galaxies with emission line fluxes of \SI{2e-16}{erg.cm^{-2}.s^{-1}} at 1.6\,\micron. Redshifts for the remaining galaxies will be measured photometrically by combining ground-based optical photometry with the \Euclid data.

\Euclid will also devote 17\% of the primary mission time to obtaining deeper observations needed for calibration, control of systematic effects, and characterising the galaxy sample from the EWS. The six \Euclid auxiliary fields (EAFs; CDFS, COSMOS, SXDS, VVDS, AEGIS, and GOODS-N, see Sect.~\ref{sec:fields} for details) each have extensive UV--NIR imaging observations and large catalogues of spectroscopic redshift measurements. Three large (10--20\,deg$^2$) Euclid Deep Fields (EDFs; EDF-North, EDF-South, and EDF-Fornax) were also selected for observations two mag deeper than the EWS. Consistent processing and photometric measurements from the \Euclid and ancillary data in the EAFs and EDFs is essential for calibrating photometric redshift measurements and quantifying biases in shape measurements from the broader EWS.  

The Cosmic DAWN Survey (DAWN) is a 59\,deg$^2$ multiwavelength imaging survey of the EDFs and EAFs with comparable depth to the \Euclid data. \textit{Spitzer}/IRAC data cover all the DAWN fields at 3.6 and 4.5 \micron\ \citep{moneti2022} and incorporate the single largest allocation of \textit{Spitzer} observing time \citep{capak2016}. The Hawaii Twenty deg$^2$ Survey \citep[H20;][]{DAWN-DR1} provides Subaru Hyper Suprime-Cam optical data and CFHT MegaCam UV data for EDF-N and EDF-F. The Hyper Suprime-Cam Subaru Strategic Survey \citep[HSC-SSP][]{Aihara2018} provides optical data for COSMOS, SXDS, VVDS, and AEGIS. Additional UV--optical data for EDF-F and EDF-S will also be provided by Vera C. Rubin Observatory \citep{ivezic2019}. The CFHT large area U-band deep survey \citep[CLAUDS;][]{Sawicki2019} and the ongoing Deep Euclid U-band Survey (DEUS) program provide additional CFHT MegaCam UV data in COSMOS, SXDS, EDF-N, and GOODS-N.

\begin{figure}
    \centering
    \includegraphics[clip, trim=0mm 1mm 2mm 5mm, width=0.5\textwidth]{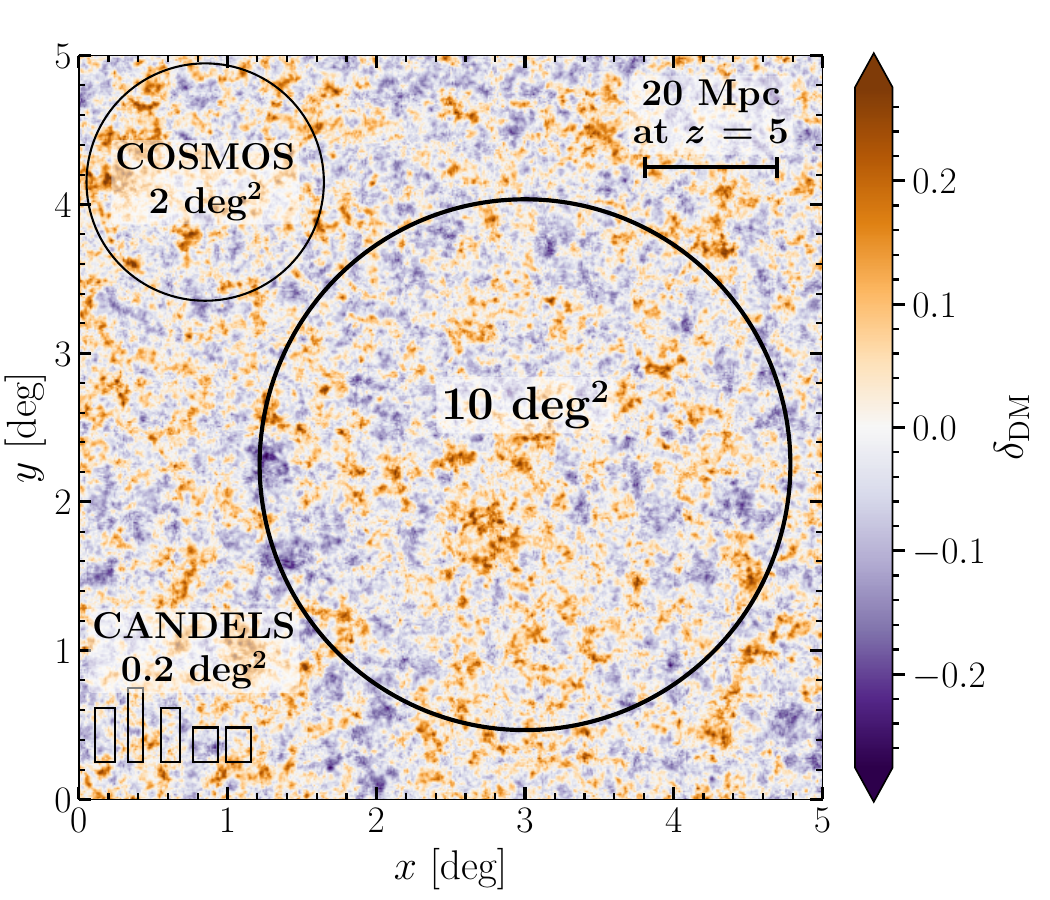}
    \caption{Illustration of the dark matter overdensity, $\delta_\mathrm{DM} \equiv \rho_\mathrm{DM} / \bar{\rho}_\mathrm{DM} - 1$, at redshifts $4.3 < z < 5.3$ in a $\ang{5} \times \ang{5}$ region from the cosmological simulation of \citet{Racz2021}. The large circle encloses a 10\,deg$^2$ region, which is equivalent to the area covered by the smallest EDF. The small rectangles and circle are comparable to the areas of the CANDELS and COSMOS surveys, respectively. The DAWN survey has both the depth and large area coverage needed to study rare overdensity peaks (orange) and cosmic voids (blue) and place strong constraints on the overall cosmological distribution of dark matter. The scale bar shows the angular size for a proper distance of 20\,Mpc at $z = 5$.}
    \label{fig:DM}
\end{figure}

The combination of depth, area, and wavelength coverage makes DAWN an excellent data set for studying galaxy evolution and robustly characterising the $z>4$ galaxy population. Deep rest-frame UV--near-IR data are critical for estimating photometric redshifts and physical parameters (e.g., stellar mass, star-formation rate) of galaxies. Robustly characterizing the $z > 4$ galaxy population requires multiwavelength data that cover observed wavelengths from the near-UV (0.3\,\micron) to near-IR (3--5\,\micron) and are deep enough to detect faint ($z_\mathrm{AB} \sim 26$) galaxies. Large fields ($\gtrsim$10\,deg$^2$) are needed to identify rare sources in the high-redshift Universe, probe cosmologically significant volumes ($\sim$1\,Gpc$^3$), and making them contiguous allows one to study the evolution of large-scale structure with cosmic time. Existing deep fields with similar depth and wavelength coverage such as the CANDELS fields \citep{Grogin2011,Koekemoer2011} and COSMOS deep field \citep{Scoville2007} are prone to small number statistics and cosmic variance due to the scale of variations in the underlying dark matter (DM) density field at high redshifts (see Fig.~\ref{fig:DM}). 

The COSMOS field has shown the power of multiwavelength data sets spanning large areas of the sky. The DAWN survey aims to expand on the legacy of COSMOS by providing value added catalogues with consistently measured \Euclid selected multiwavelength photometry and physical properties in the EDFs and EAFs; an area 30$\times$ larger than the COSMOS field. By including \textit{Spitzer}/IRAC data needed to probe the rest-frame optical at $z > 4$, the DAWN catalogues are optimized for high-redshift and galaxy evolution science and are thus complimentary to the official \Euclid catalogues. This paper describes the fields, observations, and science goals of the DAWN survey. A companion paper \citep{DAWN-DR1} provides the first DAWN survey catalogue of the pre-launch data in EDF-N and EDF-F. Future DAWN data releases (including EDF-S and the EAFs) will follow each of the \Euclid data releases. 

Section~\ref{sec:fields} describes the fields covered by the DAWN survey. Sections~\ref{sec:image-data}~and~\ref{sec:spectroscopy} summarise the broad-band imaging and spectroscopic observations (respectively) in each field. The main science goals of the survey are discussed in Section~\ref{sec:science-goals} and Section~\ref{sec:summary} provides a summary of the survey.

\begin{table}[t!]
\caption{The center coordinates (J2000), area coverage, and foreground dust reddening $E(B-V)$ of the Euclid Deep and Auxiliary fields.}
\smallskip
\label{table:fields}
\smallskip
\begin{center}
\begin{threeparttable}   
\begin{tabular}{|l|c|c|c|c|}
  \hline
  & & & & \\[-9pt]
  Field & RA & Dec & Area & $E(B\!-\!V)$\tnote{*}\\
   &  & & [deg$^2$] & [mag]\\
  & & & & \\[-9pt]
  \hline
  & & & & \\[-9pt]
  EDF-N   & 17:58:55.9 & +66:01:03.7 & 20 & 0.045 \\
  EDF-F   & 03:31:43.6 & $-$28:05:18.6 & 10 & 0.009\\
  EDF-S   & 04:04:57.8 & $-$48:25:22.8 & 23 & 0.017\\
  \hline
  COSMOS  & 10:00:28.6 & +02:12:36.0 & 2 & 0.016\\
  SXDS    & 02:18:00.0 & $-$05:00:00.0 & 2 & 0.013\\
  AEGIS   & 14:19:18.5 & +52:49:12.0 & 1 & 0.013\\
  VVDS    & 02:26:00.0 & $-$04:30:00.0 & 0.5 & 0.018\\
  CDFS    & 03:32:28.1 & $-$27:48:36.0 & 0.5 & 0.009\\
  GOODS-N & 12:37:00.0 & +62:15:00.0 & 0.5 & 0.011\\
  \hline
\end{tabular}
\end{threeparttable}
\begin{tablenotes}
    \item[*]  The provided $E(B\!-\!V)$ values are evaluated at the field center from \citet{PlanckDust}.
\end{tablenotes}
\end{center}
\end{table}

\begin{figure*}
    \centering
    \includegraphics[width=\textwidth]{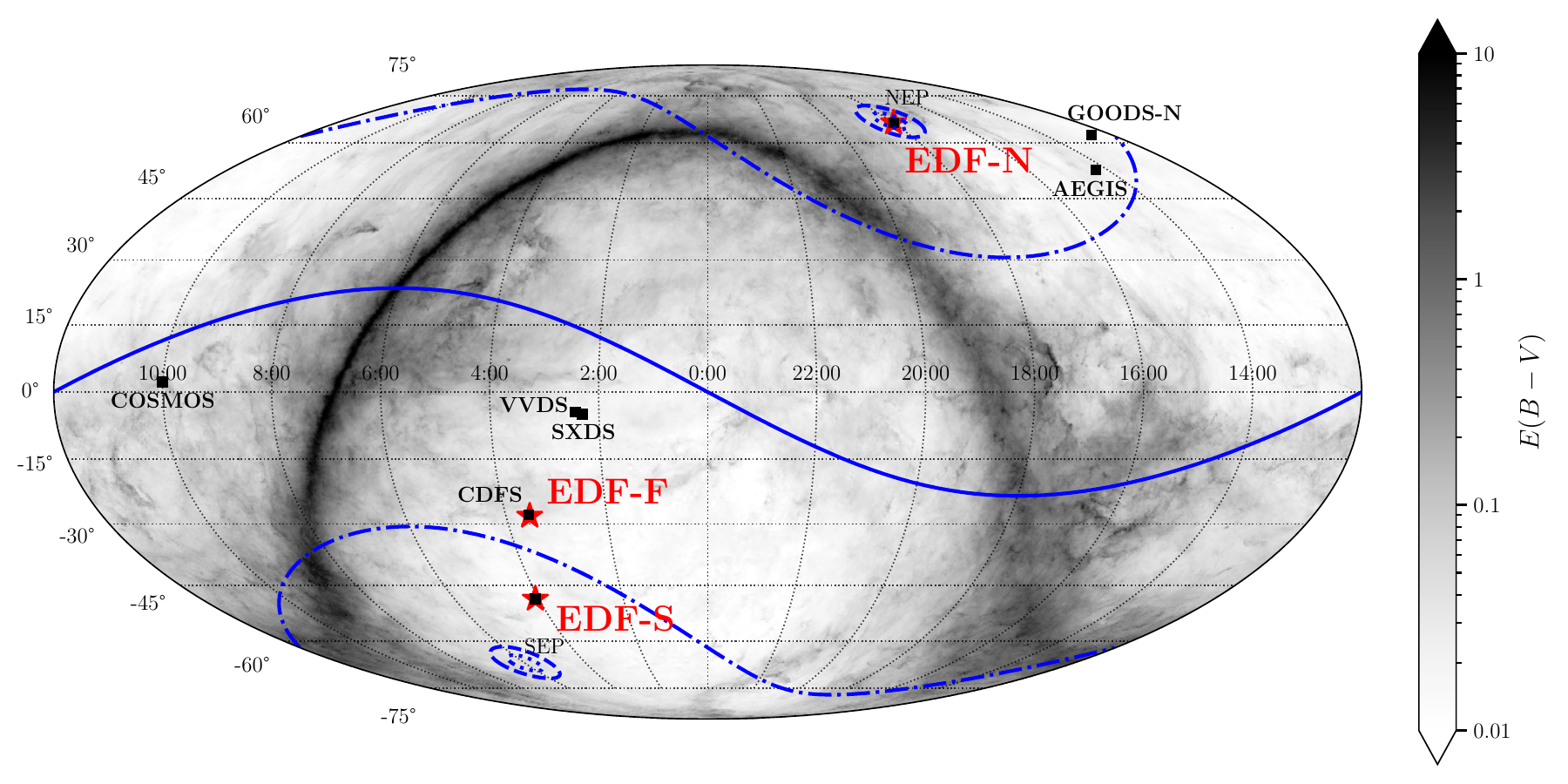}
    \caption{A Mollweide projection of the Galactic foreground reddening -- $E(B-V)$ -- on the celestial sphere from the \citet{PlanckDust} generated using the \texttt{dustmaps} Python package \citep{dustmaps}. The coordinates of each of the EAFs are indicated with black squares and the EDFs are highlighted with red stars. The solid blue line shows the ecliptic. The dotted blue lines indicate continuous viewing zone (CVZ) for the \Euclid mission at $\pm$\ang{87.5}. The dot-dashed blue lines indicate ecliptic latitudes of $\pm$\ang{54} which is the expected boundary for the CVZ of the \textit{Roman} space telescope \citep{foley2018}. The dashed blue lines show the CVZ for JWST at $\pm$\ang{85}. The locations of the north and south ecliptic poles are indicated with the text NEP and SEP, respectively.}
    \label{fig:skymap}
\end{figure*}

\begin{table*}[t!]
\caption{Expected photometric depths for each of the fields covered by the DAWN survey. Unless noted below, depths are quoted as 5$\sigma$ limiting magnitudes (AB) in \ang{;;2} empty apertures.}
\smallskip
\label{table:depths}
\smallskip
\begin{center}
\begin{threeparttable}
\begin{tabular}{|c|cccccccc|}
  \hline
  & & & & & & & & \\[-9pt]
  Instrument / Band & EDF-N & EDF-F & EDF-S & COSMOS & SXDS & VVDS & AEGIS & GOODS-N \\
  & & & & & & & & \\[-9pt]
  \hline
  & & & & & & & & \\[-9pt]
  CFHT MegaCam / $u$ & 26.4 & 26.4 & -- & 27.7 & 27.6 & 27.4 & 27.0 & 26.6 \\
  Subaru HSC / $g$   & 27.5 & 27.5 & -- & 28.1 & 28.1 & 27.5 & 26.5 & -- \\
  Subaru HSC / $r$   & 27.5 & 27.5 & -- & 27.8 & 27.8 & 27.1 & 26.1 & 27.8 \\
  Subaru HSC / $i$   & 27.0 & 27.0 & -- & 27.6 & 27.6 & 26.8 & 25.9 & 24.9\\
  Subaru HSC / $z$   & 26.5 & 26.5 & -- & 27.2 & 27.2 & 26.3 & 25.1 & 24.5\\
  Subaru HSC / $y$   & 25.1 & --   & -- & 26.5 & 26.5 & 25.3 & 24.4 & 23.8\\
  \hline
  & & & & & & & & \\[-9pt] 
  \textit{Spitzer} IRAC / [3.6\,$\mu$m] & 24.8 & 24.8 & 23.9 & 25.3 & 25.3 & 24.6 & 24.2 & 25.3\\
  \textit{Spitzer} IRAC / [4.5\,$\mu$m] & 24.7 & 24.7 & 23.8 & 25.3 & 25.1 & 24.4 & 24.2 & 25.1\\
  \hline
  & \multicolumn{3}{c|}{} & \multicolumn{3}{c|}{} & \multicolumn{2}{c|}{} \\[-9pt]
  \Euclid VIS / \IE  & \multicolumn{3}{c|}{28.2} & \multicolumn{3}{c|}{27.95} & \multicolumn{2}{c|}{27.7} \\
  \Euclid NISP / \YE & \multicolumn{3}{c|}{26.3} & \multicolumn{3}{c|}{26.05} & \multicolumn{2}{c|}{25.8} \\
  \Euclid NISP / \JE & \multicolumn{3}{c|}{26.5} & \multicolumn{3}{c|}{26.25} & \multicolumn{2}{c|}{26.0} \\
  \Euclid NISP / \HE & \multicolumn{3}{c|}{26.4} & \multicolumn{3}{c|}{26.15} & \multicolumn{2}{c|}{25.9} \\
  \hline
  & & \multicolumn{5}{|c|}{} & & \\[-9pt]
  \textit{Rubin} / $u$ & -- & \multicolumn{5}{|c|}{26.8} & -- & -- \\
  \textit{Rubin} / $g$ & -- & \multicolumn{5}{|c|}{28.4} & -- & -- \\
  \textit{Rubin} / $r$ & -- & \multicolumn{5}{|c|}{28.5} & -- & -- \\
  \textit{Rubin} / $i$ & -- & \multicolumn{5}{|c|}{28.3} & -- & -- \\
  \textit{Rubin} / $z$ & -- & \multicolumn{5}{|c|}{28.0} & -- & -- \\
  \textit{Rubin} / $y$ & -- & \multicolumn{5}{|c|}{26.8} & -- & -- \\
  \hline
\end{tabular}
\begin{tablenotes}
Notes: The \textit{Spitzer} IRAC depths are average values measured from the image data from \citet{moneti2022}. \Euclid depths are those expected for point sources by the end of the mission. Expected depths for Rubin data from \citet{foley2018}. The $u$ band depths in SXDS and VVDS are from \citet{Sawicki2019} and depths in AEGIS are from \citet{Gwyn2012}. Subaru HSC and CFHT MegaCam depths for COSMOS are from \citet{COSMOS2020}. HSC depths for SXDS, VVDS, and AEGIS are the average over each field provided by \citet{SSP-DR3}. 
\end{tablenotes}
\end{threeparttable}
\end{center}
\end{table*}

\section{Survey fields}\label{sec:fields}

The DAWN survey covers each of the EDFs and EAFs. The center coordinates and area of each field are presented in Table~\ref{table:fields}, and Fig.~\ref{fig:skymap} shows their positions on an all-sky map. The following subsections give a brief summary of the EDFs and EAFs. Consult the \Euclid overview paper \citep{EuclidOverview} for more details.

\subsection{Euclid Deep Fields}
The deepest observations from the \Euclid mission focus on three fields: Euclid Deep Field North (EDF-N), Euclid Deep Field Fornax (EDF-F), and Euclid Deep Field South (EDF-S). The locations of EDF-N and EDF-S were strategically chosen to maximise their observability throughout the duration of the Euclid mission. Each field will receive \Euclid imaging observations that are 2\,mag deeper than the EWS, which is essential for calibration of the bias introduced by noise in weak lensing measurements. The depth and area (see Table~\ref{table:fields}~\&~\ref{table:depths}) of the EDFs make them the primary fields for galaxy evolution and high-redshift science from the \Euclid mission.

EDF-N covers a 20\,deg$^2$ circular region centered on the North Ecliptic Pole (NEP) in the constellation Draco. Due to its proximity to the ecliptic pole, EDF-N has perennial visibility by \Euclid allowing for regularly repeated observations throughout the mission. EDF-F is a 10\,deg$^{2}$ circular region in the constellation Fornax. EDF-S covers a 23\,deg$^2$ elongated region in the southern constellation of Horologium.

\subsection{Euclid Auxiliary Fields}
The EAFs focus on well studied galaxy deep fields (COSMOS, AEGIS, SXDS, VVDS, CDFS, and GOODS-N) and will receive observations from \Euclid that are four to five times deeper than the EWS. Due to the large amount of existing multiwavelength data in these fields, the EAFs will be essential for calibrating photometric redshift estimates and the effect of colour gradients in galaxies on shear measurements for the EWS.

The COSMOS EAF \citep{Scoville2007} has deep multiwavelength observations from X-ray to radio and an extensive spectroscopic database. The COSMOS2020 catalogue \citep{COSMOS2020} provides the most up-to-date compilation of photometry, photo-$z$s, and stellar mass estimates. The Chandra Deep Field South (CDFS) EAF lies within EDF-F and contains one of the deepest observations from by the \textit{Chandra} X-ray observatory along with complementary observations from UV to radio. Two EAFs lie within the XMM Large Scale Structure (XMM-LSS) Survey field \citep{clerc2014}: the Subaru/XMM-Newton Deep Survey \citep[SXDS;][]{Sekiguchi2004} and the VIMOS VLT Deep Survey \citep[VVDS;][]{lefevre2005}. The SXDS EAF overlaps with the UKIRT Infrared Deep Sky Survey \citep[UKIDSS; ][]{Lawrence2007} UDS field which has extensive spectroscopic coverage from UDSz program \citep{Bradshaw2013,McLure2013,Maltby2016}. The VVDS EAF focuses on the VVDS-02h field which includes a total area of 0.61\,deg$^2$ from the VVDS Deep survey and 512 arcmin$^2$ from the VVDS Ultra-Deep Survey \citep{VVDS2005,VVDS2013}. The Great Observatories Origins Deep Survey North \citep[GOODS-N;][]{dickinson2003} EAF is centered on the Hubble Deep Field which contains some of the deepest observations obtained by HST along with complementary observations from the ground and space.
The All-Wavelength Extended Groth Strip International Survey \citep[AEGIS; ][]{davis2007} EAF is a multiwavelength survey field with low Galactic extinction and observations covering X-ray to radio wavelengths. AEGIS, CDFS, COSMOS, GOODS-N, and SXDS were all targeted as part of the Cosmic Assembly Near-infrared Deep Extragalactic Legacy Survey \citep[CANDELS;][0.2 deg$^2$ in total]{Grogin2011} and have also been observed by JWST through the CEERS \citep[100 arcmin$^2$]{CEERS2023}, JADES \citep[45 arcmin$^2$]{JADES2023}, PRIMER \citep[378 arcmin$^2$]{Dunlop2021}, NGDEEP \citep[10 arcmin$^2$]{NGDEEP2023}, and COSMOS-Web \citep[0.54 deg$^2$]{Casey2023} surveys.

\begin{figure*}
    \centering
    \includegraphics[width=\textwidth]{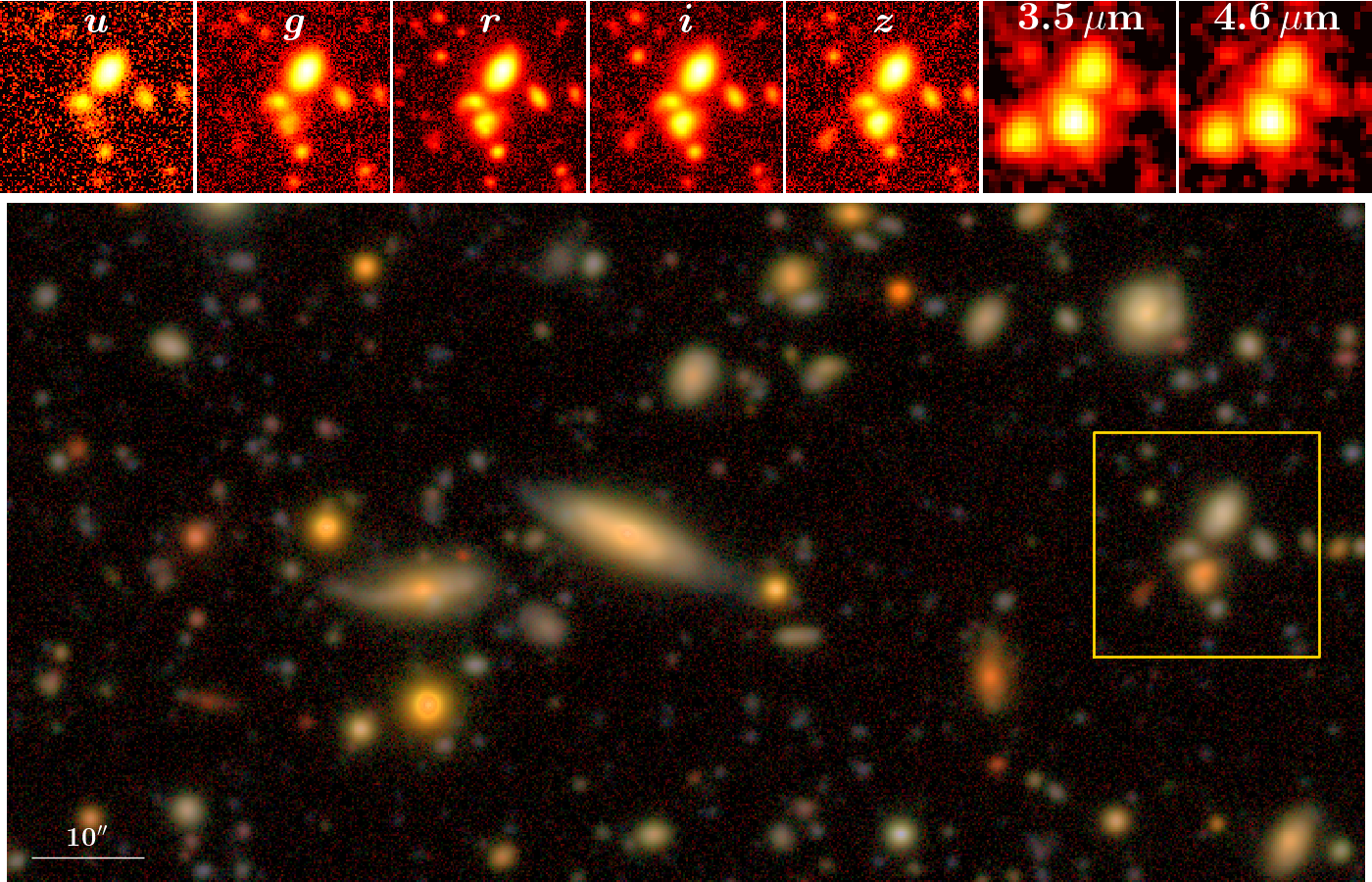}
    \caption{An example of the image data quality from the DAWN MegaCam, HSC, IRAC programs in EDF-N. The top panel shows a $riz$ colour image (\ang{;2;}\,$\times$\,\ang{;1;}) and the bottom row shows a zoom-in (\ang{;;20}\,$\times$\,\ang{;;20}) of the highlighted region in each band of the DAWN data. }
    \label{fig:image-data}
\end{figure*}

\section{Image data}\label{sec:image-data}

\begin{figure*}
    \centering
    \includegraphics[clip, trim=1mm 1mm 2mm 2mm, width=\textwidth]{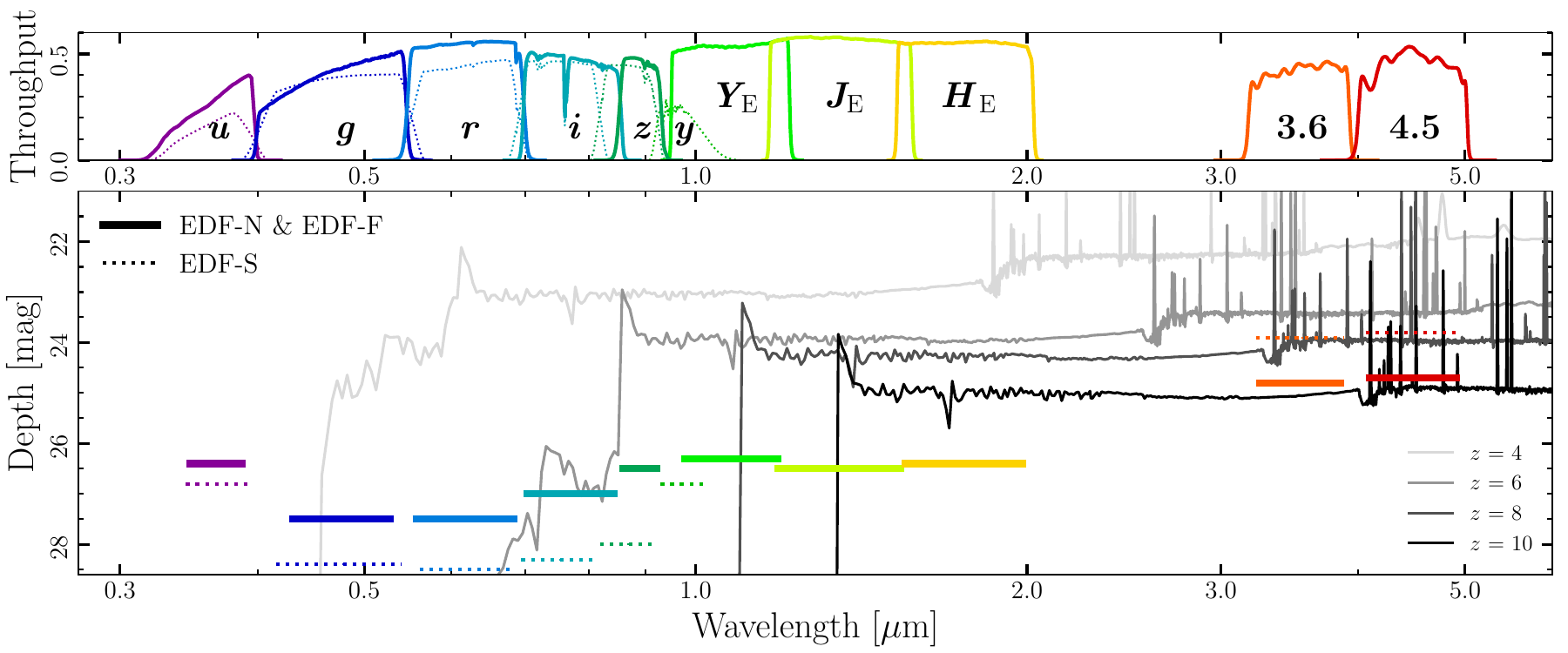}
    \caption{\emph{Top panel}: Throughput curves for the photometric bands in the DAWN survey. \emph{Bottom panel}: Photometric sensitivity limits for the EDFs at 5$\sigma$ with HSC, MegaCam, IRAC, \Euclid and \textit{Rubin} (see Tab.~\ref{table:depths} for details). The solid coloured lines show throughputs and depths for EDF-N and EDF-F data from MegaCam, HSC, and IRAC. The dotted lines show the same for EDF-S coverage from Rubin (expected) and IRAC. \Euclid NISP depths are shown with solid lines for all three fields since they are identical. The greyscale lines show \texttt{FSPS} \citep{FSPS-1,FSPS-2} model spectra for galaxies at a range of redshifts (see legend). The model spectra are normalized to stellar masses of $M^*$ galaxies at the respective redshift based on \citet{WeaverSMF} and \citet{Stefanon2021} which have values of $\log M_\star = 10.65, 10.24, 10.0, 9.5$ for redshifts $z = 4,6,8,10$; respectively. All models assume a \citet{Chabrier2003} IMF, a delayed exponential star-formation history with $\tau$ = 2\,Gyr, Solar metalicity, and \citet{Calzetti2000} dust with $A_\mathrm{V} = 0.2$. \label{fig:depths}}
\end{figure*}

The core data set of the DAWN Survey is deep space-based near-IR imaging in the \Euclid EDFs and EAFs coupled with deep wide area \textit{Spitzer}/IRAC data. Figure~\ref{fig:image-data} shows an example of the image data in EDF-N. Figures~\ref{fig:deep_fields} and \ref{fig:EAFs} show the footprints of the UV--near-IR image data in the EDFs and EAFs, respectively. Table~\ref{table:depths} provides limiting magnitudes for each band and field and Fig.~\ref{fig:depths} compares the depths with model spectra of galaxies at a range of redshifts. The rest of this section summarises the UV--mid-IR imaging observations and archival data that comprise the DAWN survey data set. 

\subsection{\Euclid Space Telescope observations}

\Euclid is a Korsch telescope with a 1.2\,m mirror providing an effective collecting area of 1\,m$^2$ \citep{racca2016} and two instruments: the \Euclid visible imager \citep[VIS;][]{VIS} has one broad visible band (\IE) that covers 530--920\,nm with a \ang{;;0.1} pixel scale; the Near-Infrared Spectrometer and Photometer \citep[NISP;][]{NISP} has three bands--\YE, \JE, and \HE-- covering 949.6--1212.3\,nm, 1167.6--1567.0\,nm, and 1521.5--2021.4\,nm \citep[respectively;][]{Schirmer2022} with a pixel scale of \ang{;;0.3}. VIS and NISP share a common field of view (FoV) of 0.53\,deg$^2$.

\Euclid observations in the EDFs and EAFs are the primary motivation for the DAWN survey. The EDFs will each receive more than 40 visits with a final depth that is 2\,mag deeper than the EWS. The EAFs will be covered by 1--4 \Euclid FoVs with varying depths listed in Table~\ref{table:depths}.

\subsection{\textit{Spitzer}/IRAC observations}

The \textit{Spitzer} Space Telescope \citep{Werner2004} Infrared Array Camera \citep[IRAC;][]{Fazio2004} is an imaging camera with four channels centered at 3.6, 4.5, 5.8, and 8.0\,\micron\ (referred to as channels 1--4, respectively). Each channel has a $\ang{;5.2;}\times\ang{;5.2;}$ FoV and a pixel scale of \ang{;;1.2}.

The DAWN survey covers the Euclid Deep Fields with uniform \textit{Spitzer}/IRAC imaging data at 3.6 and 4.5 \micron\ from two dedicated programs: the \Euclid/\textit{WFIRST} \textit{Spitzer} Legacy Survey \citep[SLS;][]{capak2016} that covers EDF-N and EDF-F; and a dedicated program to observe EDF-S \citep{Scarlata2019}. The dedicated observations of the deep fields are supplemented by previous observations extracted from the archive. Observations of the EAFs are also compiled from the archive. All available data in the \Euclid fields have been uniformly reduced and are described in detail in \citet{moneti2022}.

\subsection{HSC optical observations}
Hyper Suprime-Cam \citep[HSC;][]{Miyazaki18} is a wide-field optical imaging camera at the prime focus of the 8.2\,m Subaru telescope on Maunakea, Hawaii. With a \ang{1.5} diameter FOV and a \ang{;;0.168} pixel scale combined with the large collecting area of Subaru, HSC is an efficient instrument for deep surveys covering large areas of the sky.

\subsubsection{H20}
The Hawaii Twenty deg$^2$ Survey (H20) covers two 10 deg$^2$ fields in the EDF-N and EDF-F with the Subaru HSC $griz$ bands. Observations of each field were obtained using a seven point flower petal pattern with one central pointing surrounded by the remaining six spread over a circle with a radius of 1.1 deg (see Fig.~\ref{fig:deep_fields}). Pointings are executed using a standard five point dither pattern with a throw of 120\arcsecond. The target exposure times are 1.1, 2.5, 4.1, and 4.8 hours per pointing for the $griz$ bands, respectively. As these observations are still ongoing, we defer a full accounting of the total number of exposures to a future publication describing the final DAWN data release.

\subsubsection{HSC-SSP}
The Hyper Suprime-Cam Subaru Strategic Program \citep[HSC-SSP;][]{Aihara2018} is a 330 night $grizy$ band imaging survey of the extragalactic sky. HSC-SSP is a three tier `wedding cake' survey composed of a 1400\,deg$^2$ wide layer, a 26\,deg$^2$ deep layer covering four fields, and a 3.5\,deg$^2$ ultra-deep layer covering COSMOS and SXDS. The extended COSMOS field and XMM-LSS, which encompasses SXDS and VVDS, were both observed as part of the deep survey. A single pointing centered on the AEGIS field was also observed to the depth of the wide survey for photo-$z$ calibration. The HSC-SSP observing strategy is discussed in \citet{Aihara2018} and details of the data reduction are described in \citet{Bosch2018} and \citet{SSP-DR1,SSP-DR2,SSP-DR3}.

\subsubsection{Archival HSC data}
Additional HSC observations were extracted from the archive where available. The HEROES \citep{HEROES} and AKARI-NEP \citep{Oi2021} programs both obtained $grizy$ observations in EDF-N. Various PI-led programs at the University of Hawaii provide supplementary data for the COSMOS \citep[see][for details]{Hu2016,tanaka2017} and GOODS-N fields.

\subsection{MegaCam ultraviolet observations}
MegaCam \citep{Boulade2003} is an optical wide-field imaging camera mounted at the prime focus of the 3.6\,m Canada France Hawaii Telescope (CFHT) on Maunakea, Hawaii. MegaCam has a \ang{1;;}\,$\times$\,\ang{1;;} FoV and a pixel scale of \ang{;;0.187}. Over its lifetime, MegaCam has had two different $u$ band filters: an original $u^*$ filter that was discontinued in 2015 due to a red leak around 5000\,\AA\ and a new bluer $u$ filter that has been in use since then \citep{Sawicki2019}.

The H20 team conducted a dedicated MegaCam $u$ band imaging survey in EDF-N and EDF-F. Observations were obtained in a $4\times4$ grid with a total area coverage of 13.7\,deg$^2$ in each field. In the EDF-N the ongoing 240-hour Deep Euclid U-band Survey (DEUS; PIs:  Arnouts and Sawicki) will reach a depth of $u=27$ AB over $\sim$10\,deg$^2$ when completed; the DEUS data taken to date were shared by the DEUS team and used in making the $u$ band mosaic for this field. The third Euclid Deep Field, EDF-S, has not been observed by MegaCam as it is not accessible from Maunakea. Planned observations of EDF-S by the Rubin observatory are discussed in Sect.~\ref{sec:lsst}.
 
A variety of programs provide MegaCam observations of the EAFs. The CFHT Large Area U-band Deep Survey \citep[CLAUDS;][]{Sawicki2019} performed dedicated observations using the $u$ band in the COSMOS EAF. CLAUDS and the MegaCam Ultra-deep Survey: $u^*$ band Imaging \citep[MUSUBI;][]{MUSUBI} survey provide $u^*$ band in XMM-LSS and COSMOS. Observations of the AEGIS EAF were obtained through the CFHT Legacy Survey (CFHTLS) \citep{Gwyn2012} in the $u^*$ band with deep imaging in the CFHTLS-D3 field covering 1\,deg$^2$ and shallower data in the larger CFHTLS-W3 field covering 7\,deg$^2$. Archival $u^*$ band observations of GOODS-N (PI: L. Cowie) were extracted from the Canadian Astronomy Data Centre\footnote{\url{http://www.cadc-ccda.hia-iha.nrc-cnrc.gc.ca/}} (CADC) and reduced in the same manner as the data obtained by the H20 team.

\subsection{\texorpdfstring{$K$}{K} band observations}

Ground-based observations in the $K$ band provide supplementary data to fill in the wavelength gap between the \Euclid NISP and \textit{Spitzer}/IRAC data. Various programs have obtained $K$ band data in the DAWN fields which we briefly summarise here and provide relevant citations where available. The UltraVISTA program \citep{McCracken2012,Moneti2023} performed $K_\textrm{s}$ band observations of the COSMOS field using VIRCAM on the 4\,m VISTA telescope at Cerro Paranal Observatory in Chile. Deep observations of 20\,deg$^2$ in EDF-S have been executed through the EDFS-Ks program (PI M. Nonino). The VIDEO \citep{jarvis2013} and VEILS \citep{VEILS2017} surveys provide VIRCAM $K_\textrm{s}$ data in XMM-LSS and CDFS. Additional $K$ band observations from the UKIRT WFCAM also exist in SXDS from UKIDSS UDS \citep{Lawrence2007}.


\subsection{Future Rubin observations}\label{sec:lsst}
The Vera Rubin Observatory will perform repeated observations over \num{18000}\,deg$^2$ to identify optical transients, study dark energy and dark matter, map the Milky Way, and characterise the population of Solar System objects through the 10 year Legacy Survey of Space and Time (LSST) survey \citep{ivezic2019}. In addition to the main LSST program, Rubin will also observe a set of deep drilling fields with more frequent visits that will ultimately have much deeper data in the $ugrizy$ bands than the main survey. Although the specific fields for the deep drilling survey have not been finalised at this time, they will likely include EDF-S, CDFS, COSMOS, and XMM-LSS. When available, these data will also be incorporated into the DAWN survey data set.

\section{Spectroscopic follow-up}\label{sec:spectroscopy}

Spectroscopic follow-up is essential for precise redshift measurements, confirming objects of interest, and photo-$z$ calibration. Extensive spectroscopic surveys have already been conducted in the EAFs and will be summarised along with the DAWN catalogue release for those regions. This section provides a brief summary of ongoing spectroscopic follow-up in the EDFs.

\subsection{Keck DEIMOS}

The H20 team in Hawaii is conducting an ongoing spectroscopic survey of high-redshift galaxies in the EDF-N and EDF-F using the Deep Extragalactic Imaging Multi-Object Spectrograph \citep[DEIMOS;][]{Faber2003} on the 10\,m Keck II telescope. The survey, so far, has focused on Lyman break galaxies \citep{Steidel1996}, also known as `dropout' galaxies, selected using observed optical colours. Specifically, the program has focused on targeting Ly-$\alpha$ emission in overdensities of $g$ band dropout galaxies at $z\sim4$. See Murphree et al. (in prep.) for a full description of the colour selection criteria, overdensity calculations, and results.

High-$z$ galaxies in protoclusters are scientifically interesting and convenient observationally since a large number of galaxies can be observed in a single mask. These observations increase the number of high-redshift spec-$z$ measurements which have immediate impact in achieving the science goals of the DAWN survey and help to improve redshift estimates derived from the photometry.

\subsection{Hobby-Eberly Telescope VIRUS IFU}

The Texas-Euclid Survey for Lyman-Alpha \citep[TESLA;][]{ChavezOrtiz2023} is a spectroscopic survey targeting a 10\,deg$^2$ region in the center of EDF-N to generate a large sample ($\sim$\num{50000}) of Ly-$\alpha$ emitting galaxies (LAEs) at redshifts $z$\,=\,2--3.5 to explore how the physical properties of LAEs correlate with emerging Ly-$\alpha$ emission. The combination of these spectra with the deep H20 imaging allows redshift identification of TESLA-identified emission lines via spectral-energy distribution (SED) fitting, and subsequent analyses including the study of the physical properties of identified LAEs.  TESLA spectroscopic data is acquired by the Visible Integral-field Replicable Unit Spectrograph \citep[VIRUS;][]{Hill2018} instrument atop the Hobby-Eberly Telescope. The VIRUS instrument has a wavelength coverage of 3500--5500\,\AA\ with a resolving power of R $\sim$ 800 making the VIRUS instrument optimal for detecting Ly-$\alpha$ emission from galaxies at redshifts $z$\,=\,1.9--3.5, as well as \ion{O}{ii}-emitting galaxies at $z <$ 0.5.

\section{Science goals}\label{sec:science-goals}

The DAWN survey's deep NIR imaging from \Euclid, complemented by matched-depth UV to IR data, facilitates a wide array of scientific analyses. Sharing similarities in wavelength coverage and depth with prior surveys like CANDELS \citep{Grogin2011} and COSMOS \citep{Scoville2007}, DAWN stands out due to its significantly larger area coverage—thirty times that of the COSMOS field. This expanded scope opens up numerous new analytical possibilities. The primary scientific goals of the DAWN survey revolve around understanding galaxy formation and evolution, particularly in the high-redshift Universe. This section outlines the core objectives for the survey team, although the dataset will support a multitude of other research endeavors as well.

\subsection{UV luminosity functions and mapping reionization}
\begin{figure*}
    \centering
    \includegraphics[clip, trim=3mm 3mm 5mm 3mm, width=\textwidth]{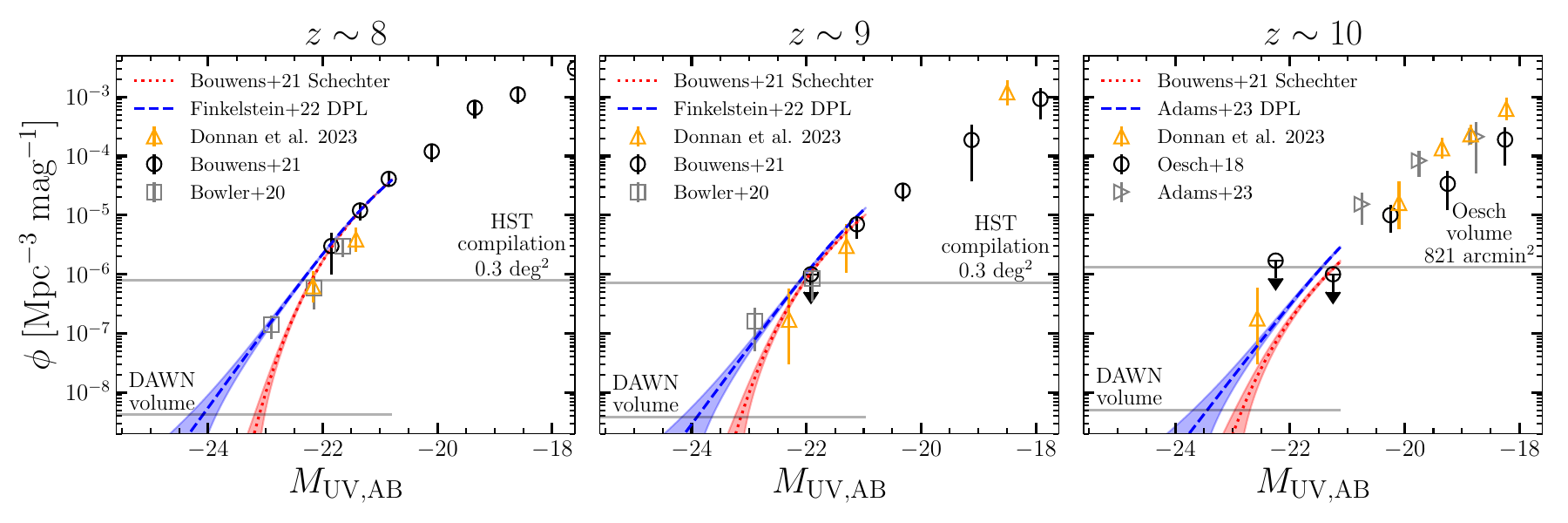}
    \caption{Measurements of the ultraviolet luminosity function $\phi$ at $z = \textrm{8--10}$ from the literature \citep{Oesch2018,Bowler2020,Bouwens2021,Finkelstein2022,Adams2023,Donnan2023}. The red and blue lines show extrapolations to the volume of the DAWN survey based on the best fitting Schechter and double power-law (DPL) functions, with the shaded regions showing associated Poisson uncertainties for the DAWN survey volume. Upper limits based on survey volume are indicated with the horizontal lines. The DAWN volume in each panel assumes an area of 59\,deg$^2$, a redshift slice of $\pm0.5$ from the redshift bin center, and a \citet{Planck2018} $\Lambda$CDM cosmology.\label{fig:uvlf}}
\end{figure*}

\begin{figure}
    \centering
    \includegraphics[clip, trim=5mm 3mm 2mm 5mm, width=0.5\textwidth]{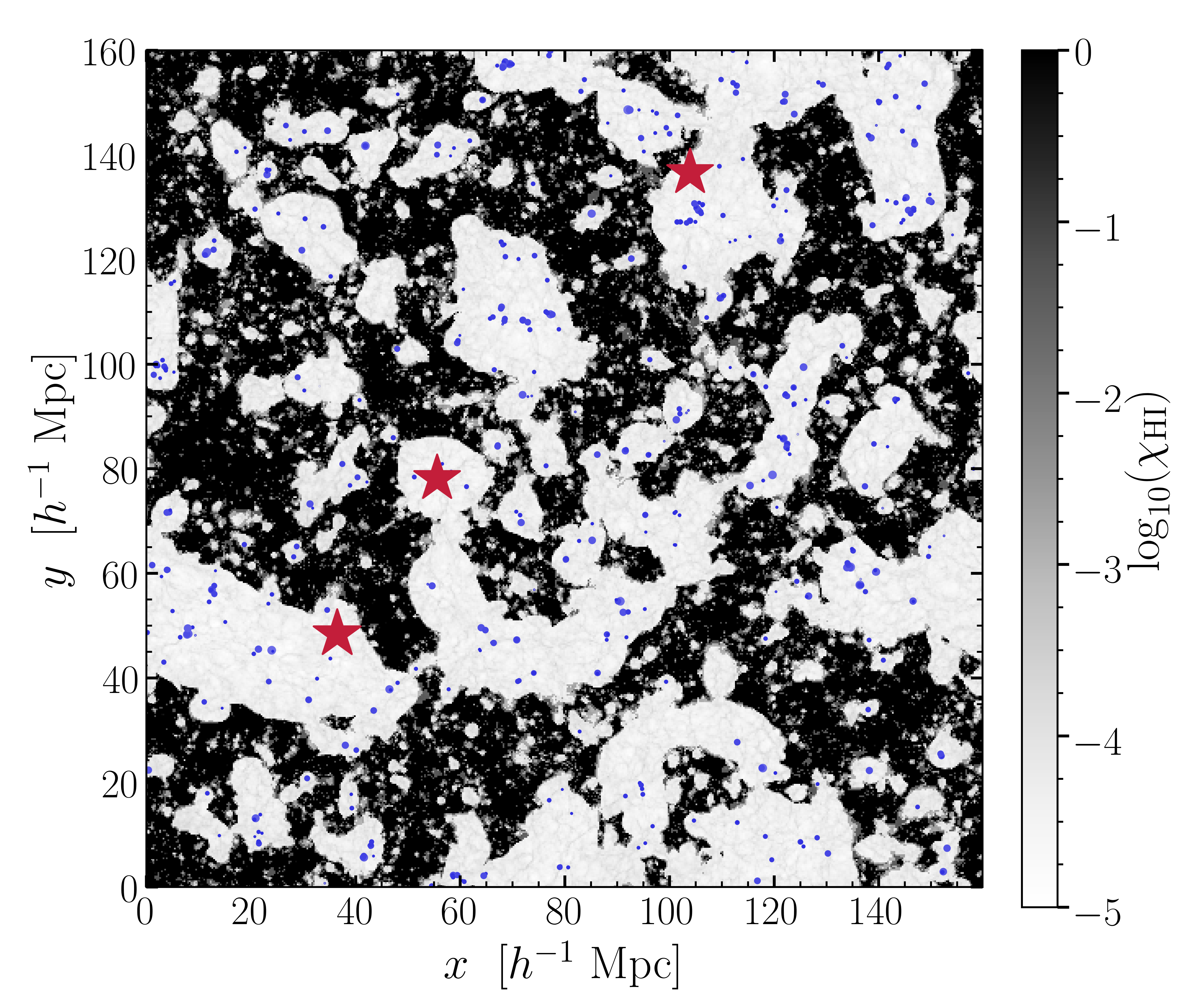}
    \caption{Large-scale spatial distribution of the neutral hydrogen fraction at $z\simeq7$ (neutral = black, white/gray=ionised) of the MHDEC (i.e. $f_\mathrm{esc}$ decreases with halo mass, $M_\mathrm{h}$) simulation presented in \citet{Hutter2023}. The neutral hydrogen fraction shown here is averaged over 3 cells along the line of sight, where each cell has a comoving depth of $312.5\,h^{-1}$kpc and the dimensionless Hubble constant $h=0.6777$. The width and height of the box are given in comoving units and correspond to \ang{1;;} at $z=7$. Blue circles depict galaxies with UV luminosities of at least $M_\mathrm{UV}\geq-18$ and red stars the brightest galaxies in the simulation box with $M_\mathrm{UV}\geq-21$.}
    \label{fig:reionization}
\end{figure}

The Epoch of Reionisation (EoR) marks the period of `cosmic dawn' in which light from the first galaxies ionised the predominately neutral intergalactic medium (IGM) allowing this radiation to stream freely throughout the Universe. Although observations of the cosmic microwave background (CMB) \citep{Planck2018}, high-$z$ quasars \citep{Bouwens2015,Robertson2015}, and Ly-$\alpha$ emitting galaxies \citep{Treu2012,Castellano2016,Kakiichi2016} constrain the end of the EoR to $z \gtrsim 6$, many important questions remain. When did the EoR begin? What is the relative contribution of rarer luminous galaxies versus more numerous faint galaxies to the budget of ionizing photons \citep[e.g.,][]{Naidu2020,Hutter2021,Hutter2023}? How does the reionization efficiency and the Ly-$\alpha$ escape fraction scale with the local density of galaxies? 

One of the primary means of addressing these questions is through measurements of the rest-frame UV luminosity function (UVLF), which describes the co-moving volume density of galaxies as a function of their UV luminosity. Integrating the UVLF not only yields an estimate of the ionizing photon budget at a given redshift but also provides constraints on the unobscured cosmic star-formation rate density \citep[see][for a review]{Madau2014}. Although deep surveys from HST and JWST have been and will continue to be instrumental for measuring the faint end of the UVLF \citep[e.g., ][]{Oesch2018,Bouwens2021,Casey2023}, larger survey areas are needed to definitively constrain the abundance of rare luminous sources on the bright end (see Fig.~\ref{fig:uvlf}). Thanks to the combination of depth and area, the DAWN survey will robustly constrain the bright end of the UVLF to absolute magnitudes of $M_\mathrm{UV} \sim -24$ at $z = \textrm{8--10}$, with sufficient depth to overlap at fainter magnitudes studied by HST \citep[e.g.,][]{Bouwens2021} and JWST surveys \citep[e.g.,][]{Casey2023, Donnan2023}.

Theoretical models predict that the most massive galaxies trace the large-scale structure (LSS) at high redshifts, with lower mass galaxies clustered around them \citep[e.g.,][]{MoAndWhite1996, Vogelsberger2014}. They also predict that reionization proceeds more efficiently in these over-dense regions \citep[e.g.,][]{Treu2012, Kakiichi2016, Castellano2016}. However, the details are still poorly understood \citep{Mason2020,Naidu2020,Larson2022,Finkelstein2022}. For example, what is the observational evidence for the association between massive galaxies and LSS at high-redshifts \citep{Hatfield2018,Harikane2022}?

Structures extending over \ang{0.25;;}--\ang{1.0;;} at $z > 3$ are rare and the existing pre Euclid Deep Fields do not cover a sufficient contiguous area to capture them. Thus, a survey like DAWN is essential to probe the full range of reionization conditions in diverse environments, probing the same variety of structures in the real Universe as those found in cosmological simulations. According to the Millennium simulation \citep{Springel2005}, the DAWN survey will find $\sim$125--250 proto-structures, i.e., progenitors of modern day $>\,5\times10^{14}\,M_\odot$ clusters, and $\sim$1250--2500 regions with densities more than three times the mean density of the Universe at $3 < z < 8$. For example, the comoving volume of the DAWN survey at $6.5 < z < 7.5$ will contain more than 100 DM halos with masses ($M_\mathrm{200}$) greater than $10^{12}\,M_\odot$; with only $5\pm2$ such massive halos expected in a survey like COSMOS \citep{Despali2016}. Figure~\ref{fig:reionization}
 shows reionization bubbles around luminous galaxies at $z \sim 7$ from the ASTREUS simulation framework \citep{Hutter2023} in a similar area as the COSMOS field.

\begin{figure*}
    \centering
    \includegraphics[clip,trim=4mm 4mm 4mm 4mm,width=\textwidth]{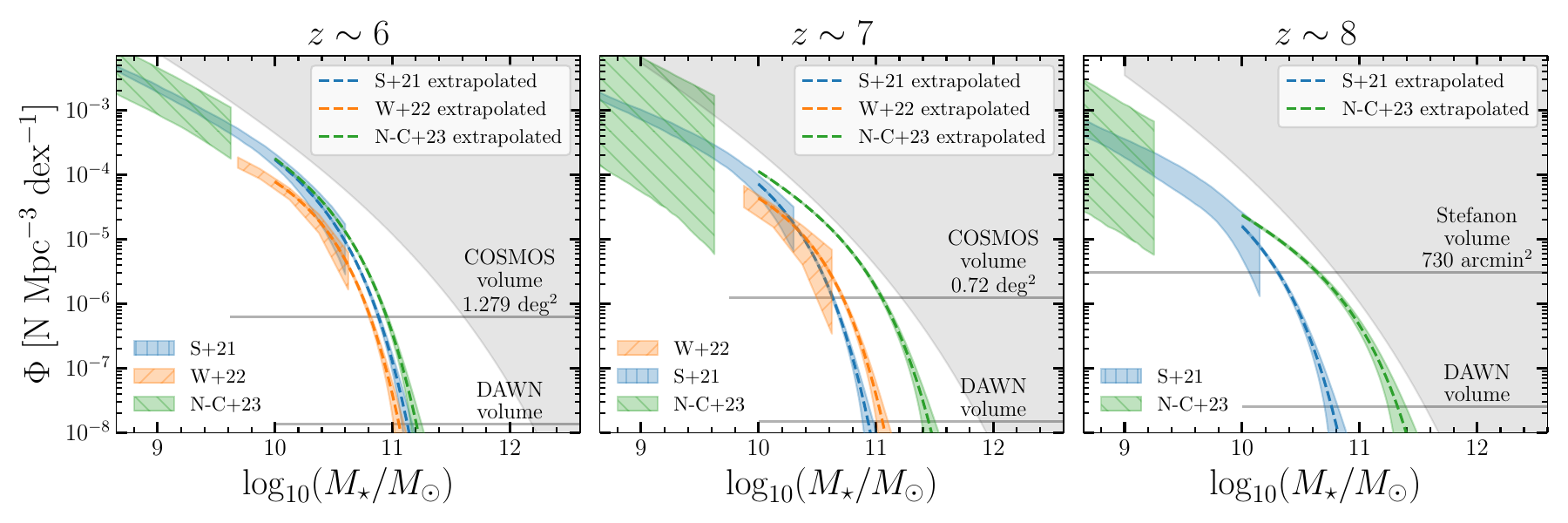}
    \caption{Stellar mass function, $\Phi$, measurements (hatched regions) from  HST \citep[S+21;][]{Stefanon2021}, the COSMOS field \citep[W+22;][]{WeaverSMF}, and JWST \citep[N-C+23;][]{Navarro-Carrera2023} in three redshift bins: $z\sim6$ (left), 7 (middle), and 8 (right). The dashed lines show extrapolations to higher masses for the DAWN survey volume (calculated in the same way as in Fig.~\ref{fig:uvlf}) assuming the corresponding literature Schechter function fits, and the accompanying shaded regions show estimated uncertainties. Upper limits for empty bins for the respective survey are indicated with horizontal lines. Note that Due to the shallower IRAC data in EDF-S (see Table~\ref{table:depths}) the $z\sim8$ bin assumes a smaller effective area of 39\,deg$^2$. The grey shaded regions indicate the theoretical upper limit of the SMF in each redshift bins based on the \citet{Despali2016} halo mass function assuming a fixed baryon fraction of 0.018 as in \citet{WeaverSMF}. 
    \label{fig:smf}
    }
\end{figure*}

\subsection{Galaxy stellar mass function}
Among the most outstanding problems in astronomy today is understanding the nature and formation mechanism of the most massive galaxies ($M_\star \gtrsim 3 \times 10^{10}\,M_\odot$) in the early Universe ($z\gtrsim2)$. These rare systems are the best candidate progenitors of local early-type galaxies with $M_\star \sim 10^{12}\,M_\odot$. Some have been shown to be surprisingly mature, having no ongoing star-formation and well-formed discs in contrast with the hierarchical formation scenario \citep[e.g.,][]{Toft2017}. In a hierarchical model, the mapping between the galaxies and host dark matter haloes is controlled by the stellar mass-halo mass relation (SMHM). Different mappings generate different merger histories as well as different star formation histories \citep[e.g.][]{Grylls2019,Fu2022}. The shape of the high-$z$ galaxy stellar mass function (SMF) is particularly valuable in this respect as it provides a benchmark for the evolution of galaxies at lower redshifts. For example, it has been shown that an abundance of massive galaxies at high redshift is more consistent with the nearly flat SFHs retrieved from galaxies at lower redshifts, which suggest that galaxies have grown much of their stellar mass at early epochs \citep{Fu2022}.

To understand the evolution of these systems one needs to address the following questions: How fast did massive galaxies build up their mass? Of those that have quenched, what mechanisms are responsible? What is the number density of different populations of galaxies at $z$\,=\, 3--8? How does environment determine their existing and future mass? 

Pursuing these questions with existing resources has already yielded surprises. While detailed follow-up of individual objects have clarified their unusual properties \citep[e.g.,][]{Toft2017}, statistical arguments as to their demographics -- namely the galaxy stellar mass function (SMF) -- have provided meaningful clues \citep{Cole2001,Adams2021,McLeod2021,Stefanon2021,COSMOS2020}. Its shape and evolution with time is sensitive not only to star-formation histories and galaxy mergers, but also to the associated physical processes thought responsible for ceasing galaxy growth such as heating of gas by accreting central supermassive black holes or by supernovae explosions, gas removal by outflows, or by a host of other proposed scenarios \citep{Dubois2013, GaborAndDave2015}.

Massive galaxies are some of the best laboratories for testing galaxy formation theories. However, a comprehensive view of massive galaxy evolution cannot come from shallow large-area surveys like SDSS or GAMA, which are restricted to $z < 0.5$, nor from deep but narrow surveys like HUDF or CANDELS, which reach $z \sim \textrm{9--10}$ but only cover a total area of $<$ 0.25\,deg$^2$ \citep[see][for a review]{Finkelstein2016}. Only a handful of high-$z$ detections have $\sim 5 \times 10^{10}\,M_\odot$ in the small cosmic volume probed by HST, with no $M_\star > 10^{11}\,M_\odot$ candidates found at $z > 5$. Even the largest contiguous HST galaxy survey, COSMOS, has not detected massive galaxy candidates beyond $z \sim 5$ \citep{WeaverSMF}. 

The DAWN survey will measure the shape and evolution of the SMF to $z \sim 8$ with directly measured stellar masses from Spitzer/IRAC. Thanks to its large area coverage, the DAWN survey will reduce Poisson uncertainties by a factor of $\sim$6 compared to the \citet{WeaverSMF} SMF from COSMOS and will provide the first robust constraints on the number density of ultra-massive galaxies ($M_\star > 10^{11}\,M_\odot$) at $z =$\ 6--8 (see Fig.~\ref{fig:smf}). 

The large area of the DAWN survey will also improve measurements of the SMF for quiescent and star-forming galaxies as well as galaxies in different environments. Colour-colour selection criteria are often used to separate star-forming and quiescent galaxies \citep{Williams2009,Arnouts2013,Ilbert2010,Davidzon2017,Gould2023}. \citet{WeaverSMF} only find 52 quiescent galaxies at $z =$\ 4.5--6.5 in 1.3\,deg$^2$ of the COSMOS field which limited their analysis of the quiescent SMF to $z < 4.5$. Assuming a similar source density, the DAWN survey will detect $\sim$2000 quiescent galaxies at $z =$\ 4.5--6.5 providing new constraints on the quiescent SMF in the early Universe. So far, measurements of the SMF as a function of environment have been limited to redshifts $z<3$ \citep{Baldry2006,Peng2010,Papovich2018,Taylor2023}. The DAWN survey will not only expand this type of analysis to higher redshifts but will also be able to explore more extreme under- or over-dense environments thanks to the large contiguous areas of the EDFs.

\subsection{Origins of large-scale structure at high-redshift}
\begin{figure}
    \centering
    \includegraphics[width=0.5\textwidth]{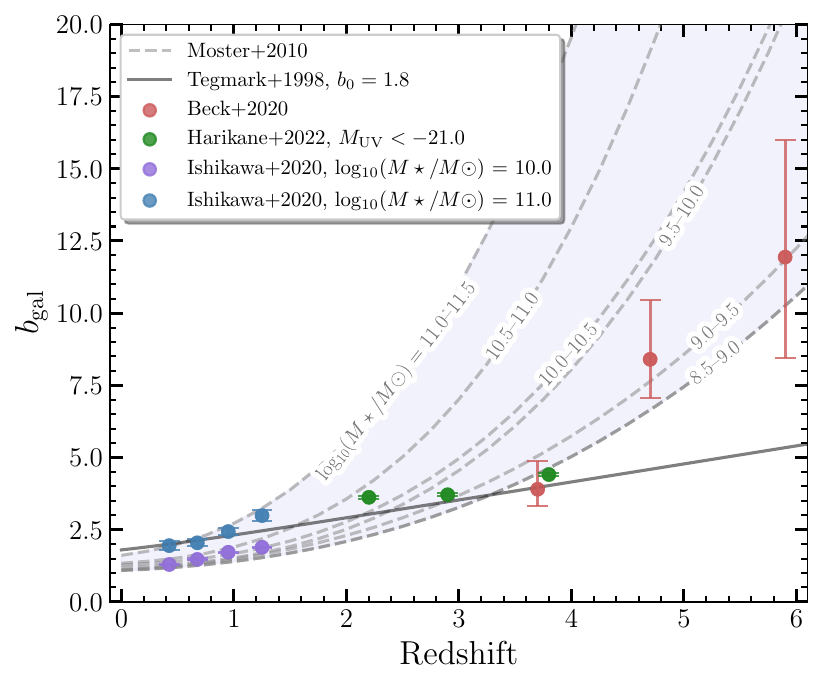}
    \caption{The redshift evolution of linear galaxy bias. The early prediction from \citet{Tegmark1998} is shown in solid grey, and more recent models separated by stellar mass from \citet{Moster2010} are shown as dashed grey curves. High-$z$ measurements for dropout-selected samples from \citet{Beck2020} and \citet{Harikane2022} are shown in red and green, respectively. Low-$z$ measurements for two stellar masses from \citet{Ishikawa2020} are shown in blue and purple. The DAWN survey will allow us to constrain bias as a function of redshift and stellar mass in the shaded region.}
    \label{fig:bias}
\end{figure}

The current standard model for cosmology, Lambda Cold Dark Matter ($\Lambda$CDM), describes the growth of structure from density fluctuations in the primordial plasma to the LSS that we observe today. The $\Lambda$CDM model has been successful in describing the power spectra of galaxies and the CMB, but recent work has found tensions \citep{abdalla2022cosmology} in two of its key parameters: $H_0$, the Hubble rate at present day \citep[e.g.,][]{Planck2018,Riess2022}, and $\sigma_8$, the amplitude of matter density fluctuations at present day \citep[e.g.,][]{Planck2018,Asgari2021,Heymans2021,Abbott2022,Amon2022}. Both tensions arise between early (CMB-based) and late (galaxy-based) methods. While $H_0$ has been measured across a wide range of redshifts, $\sigma_8$ remains unconstrained at $z$\,$\sim$\,3--7. 

The DAWN data set will allow us to measure galaxy clustering at these high redshifts. As galaxies are a biased tracer of the underlying matter distribution, we will assume a linear galaxy bias $b_\textrm{gal}$. This galaxy bias is similarly unconstrained at $z$\,$\sim$\,3--7 and is degenerate with $\sigma_8$ in two-point clustering measurements. At high redshift, we can break this degeneracy by cross-correlating galaxy clustering measurements with CMB lensing, or by combining galaxy bias measurements with higher order statistics \citep[e.g.][]{Repp2022}. Using HSC data from SSP and the University of Hawaii \citep[SSP+UH;][]{tanaka2017} in the $\sim 2$ deg$^2$ COSMOS field, \citet{Beck2020} measured linear galaxy bias out to $z \sim 6$ for $g$, $r$, and $i$ band dropouts. We will extend this analysis to the full 59 deg$^2$ of the DAWN survey and use its robust photometric redshifts to select our galaxy samples, which should reduce the uncertainties by more than a factor of three. By cross-correlating our galaxy clustering measurements with CMB lensing \citep[e.g.][]{Planck2018}, we can constrain $\sigma_8$ at high-redshift and compare our result with constraints from previous work (Murphree et al. in prep).

The total volume of the DAWN survey out to $z \sim$ 7 will be about 3.8\,Gpc$^3$. This volume is large enough to include several extreme density fluctuations. By applying advanced statistical techniques such as sufficient statistics and indicator functions, we can double this volume and make robust estimates of cosmological parameters \citep{Wolk2015,Repp2022}. The DAWN data set will allow us to constrain parameters (see Fig.~\ref{fig:bias}) well beyond the $z \sim 2.5$ forecasted for the photometric \Euclid data set \citep{EuclidXV2022}.

\section{Summary}\label{sec:summary}

The DAWN survey is a deep multiwavelength imaging survey that covers $\sim$59\,deg$^2$ in the Euclid Deep and Auxiliary fields. Deep space-based NIR image data from \Euclid NISP and \textit{Spitzer}/IRAC form the core of the DAWN data set along with complementary ground-based UV--optical data from CFHT/MegaCam, Subaru/HSC, and future observations from the Vera Rubin Observatory. The DAWN survey aims to provide catalogues for the widest area collection of extragalactic deep fields with consistent data reduction. These data will be essential for photometric redshift calibration of the Euclid Wide Survey and will be the reference data set for extragalactic studies in the Euclid Deep and Auxiliary fields.

The DAWN survey is designed to address a variety of science goals including measuring the galaxy stellar mass function to $z=8$, mapping reionization, studying the formation of large-scale structure, and characterising the first quenched galaxies. Data collection for the survey is ongoing with the ground-based component planned to be completed in the next few years and the \Euclid observations scheduled to be completed by 2030. A companion paper, \citet{DAWN-DR1}, will provide the first DAWN data release of photometric catalogues and galaxy physical parameters in EDF-N and EDF-F. Future publications will present subsequent data releases of image data and catalogues that include all of the DAWN fields. The DAWN data have immediate value for studying galaxy formation and evolution in the high-redshift Universe and that legacy value will only increase as the survey is completed.

%
%

\begin{acknowledgements}
  The Cosmic Dawn Center is funded by the Danish National Research Foundation under grant DNRF140.

  \AckEC
\end{acknowledgements}

%
%

\bibliography{references}

%
%

\begin{appendix}
  \onecolumn 
  
\section{Image data footprints}
\begin{figure*}[ht]
    \centering
    \includegraphics[clip, trim=4mm 0mm 5mm 0mm, width=\textwidth]{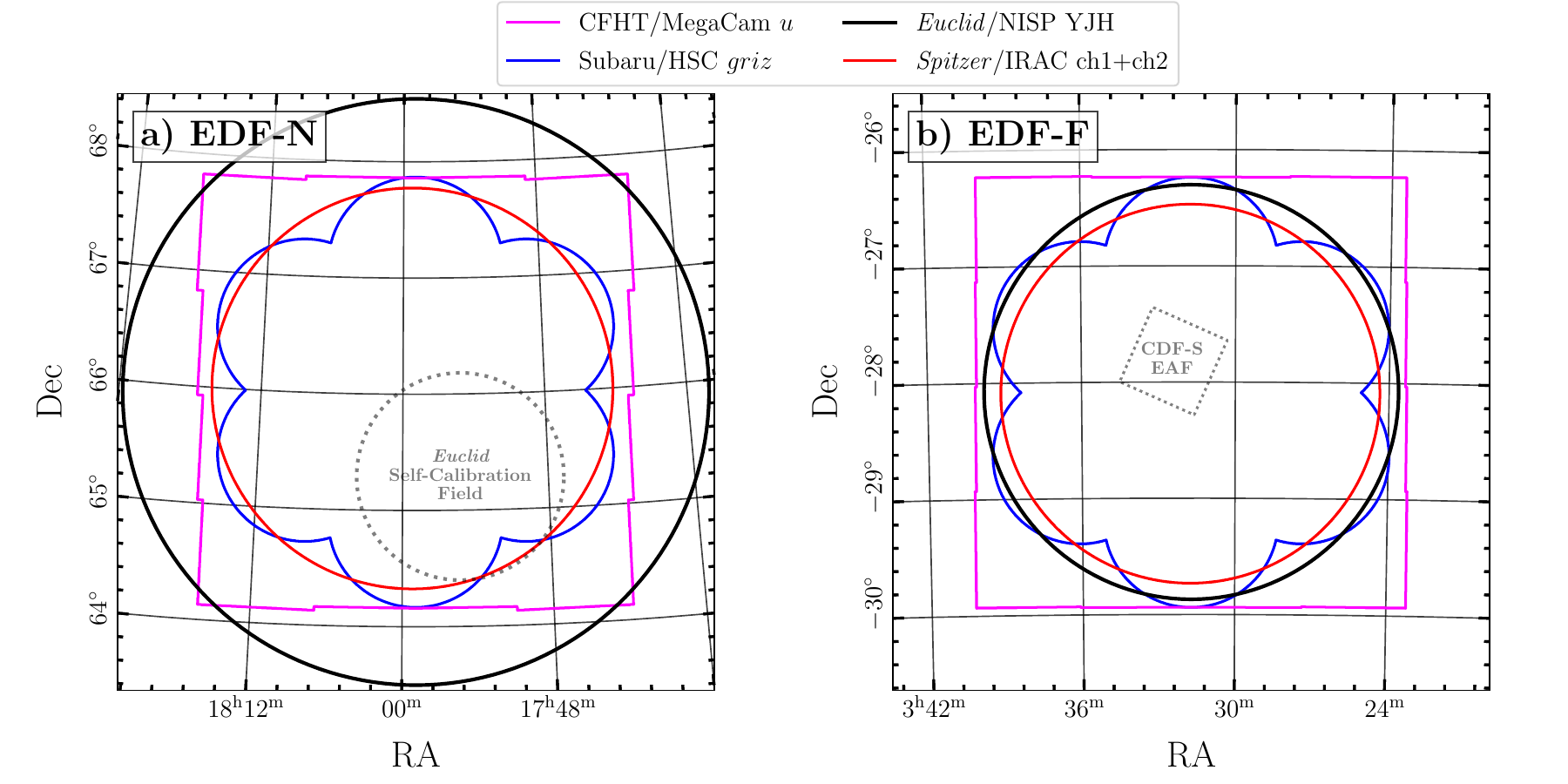}
    \includegraphics[clip, trim=5mm 3mm 5mm 3mm, width=\textwidth]{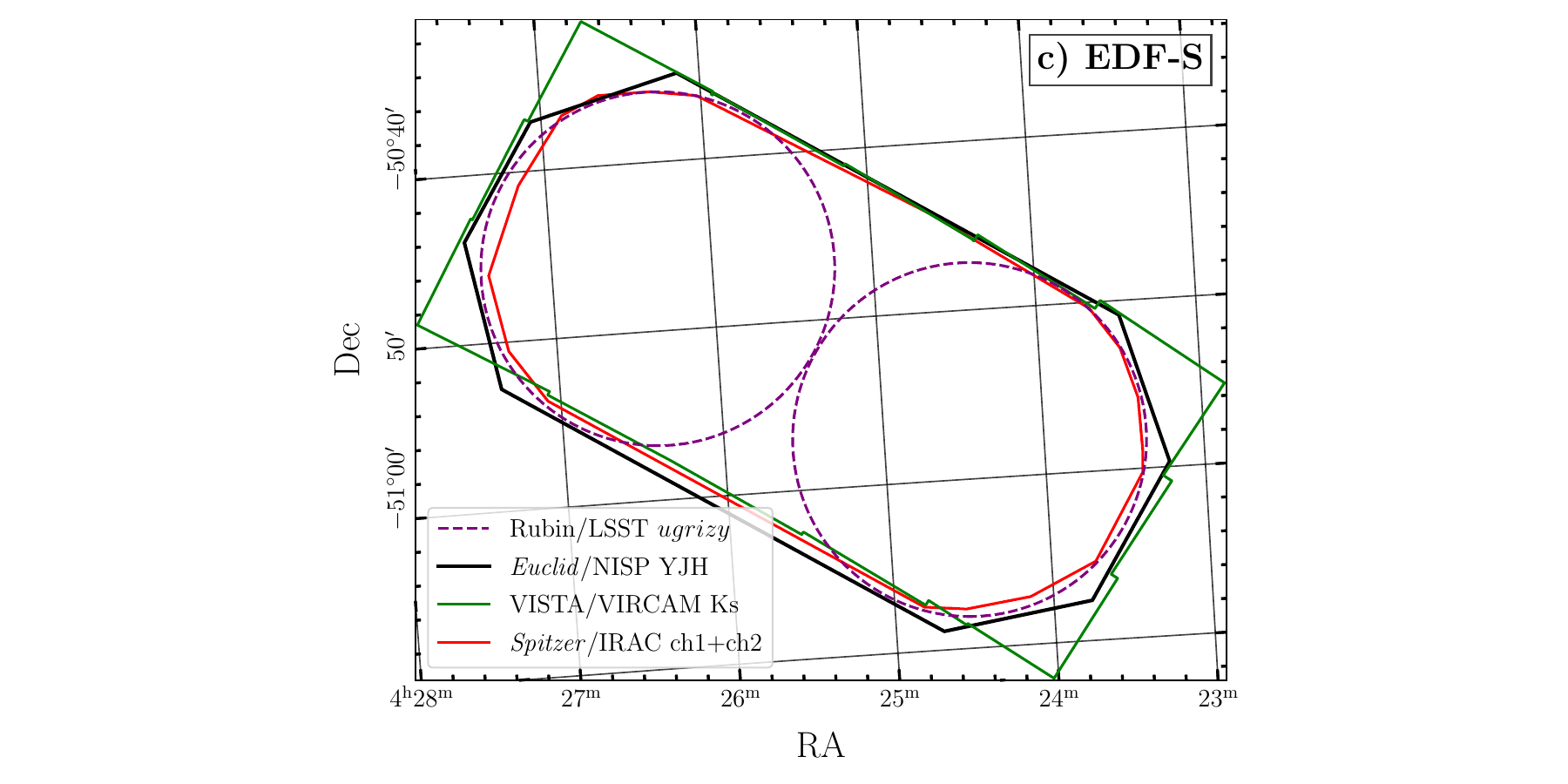}
    \caption{Footprints of the DAWN survey image data in the Euclid Deep Fields. Note that the Rubin/LSST footprint in EDF-S shows the expected pointing pattern for the EDF-S Deep Drilling Field and may change once the survey begins \citep{ivezic2019}.}
    \label{fig:deep_fields}
\end{figure*}

\begin{figure*}[ht]
    \centering
    \includegraphics[clip, trim=3mm 5mm 3mm 4mm, width=\textwidth]{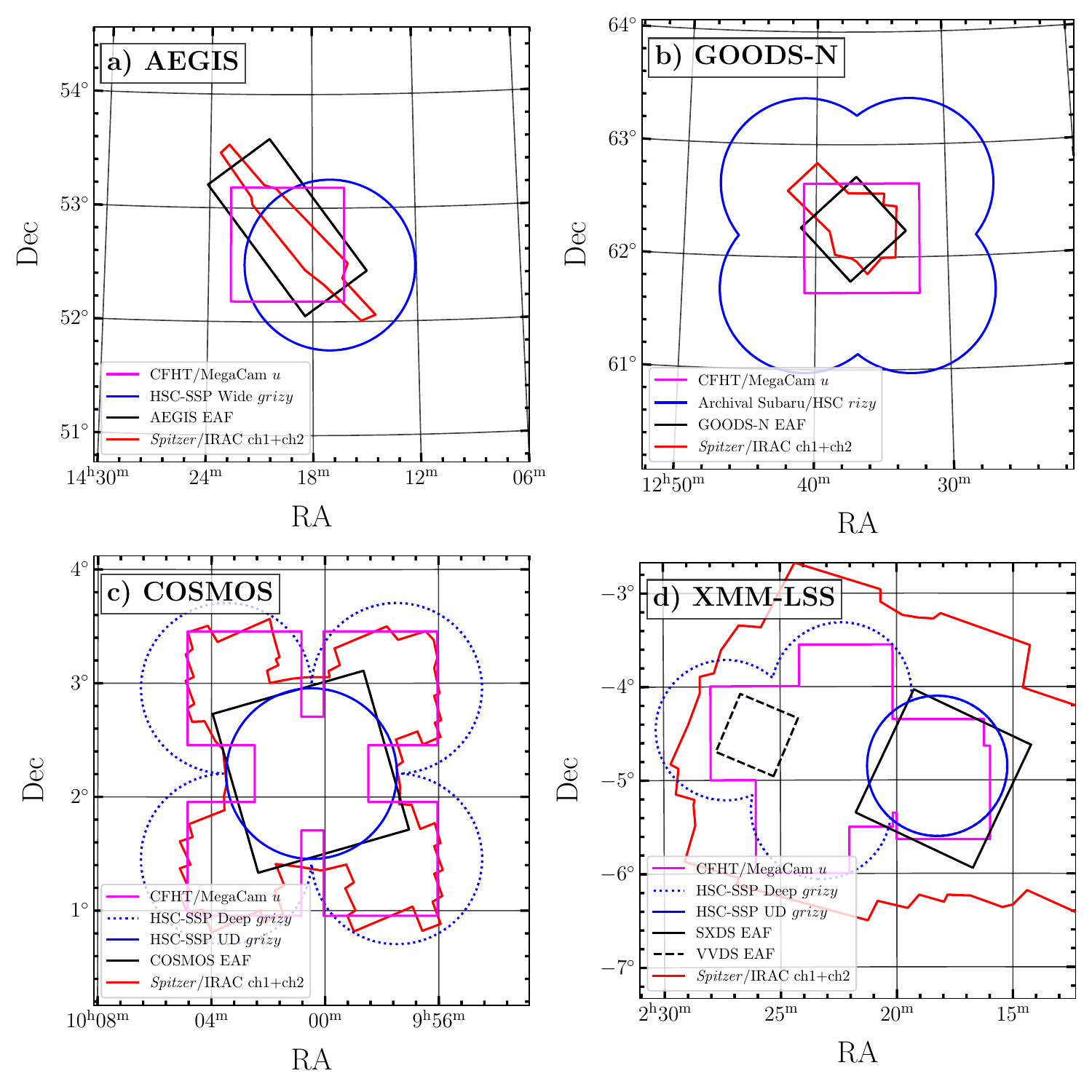}
    \caption{Footprints of the DAWN survey image data in the EAFs. The coloured lines show the extent of the image data indicated in the legend. Note that the EAF footprints are approximations based on the \Euclid survey plan. Updated footprints that to reflect the actual \Euclid coverage in the EAFs will be provided along with subsequent DAWN data releases.}
    \label{fig:EAFs}
\end{figure*}

Figures~\ref{fig:deep_fields} and \ref{fig:EAFs} show footprints of the image data compiled by the DAWN survey in the Euclid Deep and Auxiliary fields.  

\end{appendix}

\label{LastPage}
\end{document}